\documentclass[twocolumn,secnumarabic,amssymb, nobibnotes, aps, longbibliography]{revtex4-1}

\usepackage{graphicx} 
\usepackage{amsmath}
\usepackage{float}
\usepackage{color}
\usepackage[english]{babel}
\usepackage{array}
\usepackage[T1]{fontenc}
\usepackage[usenames,dvipsnames]{xcolor}
\usepackage{setspace}
\usepackage{hyperref}
\begin{document}
\title{Elastocapillary Levelling of Thin Viscous Films on Soft Substrates}
\author{Marco Rivetti$\,^1$, Vincent Bertin$\,^2$, Thomas Salez$\,^{2,3}$, Chung-Yuen Hui$\,^4$, Christine Linne$\,^1$, Maxence Arutkin$\,^2$, Haibin Wu$\,^4$, Elie Rapha\"el$\,^2$, and Oliver B\"aumchen$\,^1$}
\email[E-mail: ]{oliver.baeumchen@ds.mpg.de}
\affiliation{
$^1$ Max Planck Institute for Dynamics and Self-Organization (MPIDS),  Am Fa{\ss}berg 17, 37077 G\"ottingen, Germany. 
$^2$ Laboratoire de Physico-Chimie Th\'eorique, UMR CNRS 7083 Gulliver, ESPCI Paris, PSL Research University, 10 rue Vauquelin, 75005 Paris, France.
$^3$ Global Station for Soft Matter, Global Institution for Collaborative Research and Education,
Hokkaido University, Sapporo, Hokkaido 060-0808, Japan.
$^4$ Dept. of Mechanical \& Aerospace Engineering, Cornell University, Ithaca, NY 14853, USA.
}
\date{\today}
\begin{abstract}
A thin liquid film with non-zero curvature at its free surface spontaneously flows to reach a flat configuration, a process  driven by Laplace pressure gradients and resisted by the liquid's viscosity. 
Inspired by recent progresses on the dynamics of liquid droplets on soft substrates, we here study the relaxation of a viscous film supported by an elastic foundation. Experiments involve thin polymer films on  elastomeric substrates, where the dynamics of the liquid-air interface is monitored using atomic force microscopy. A theoretical model that describes the coupled evolution of the solid-liquid and the liquid-air interfaces is also provided. In this soft-levelling configuration, Laplace pressure gradients not only drive the flow, but they also induce elastic deformations on the substrate that affect the flow and the shape of the liquid-air interface itself. This process represents an original example of elastocapillarity that is not mediated by the presence of a contact line. We discuss the impact of the elastic contribution on the levelling dynamics and show the departure from the classical self-similarities and power laws observed for capillary levelling on rigid substrates. 
\end{abstract}
\maketitle

\section{Introduction}
\begin{figure*}
\includegraphics[width=2\columnwidth]{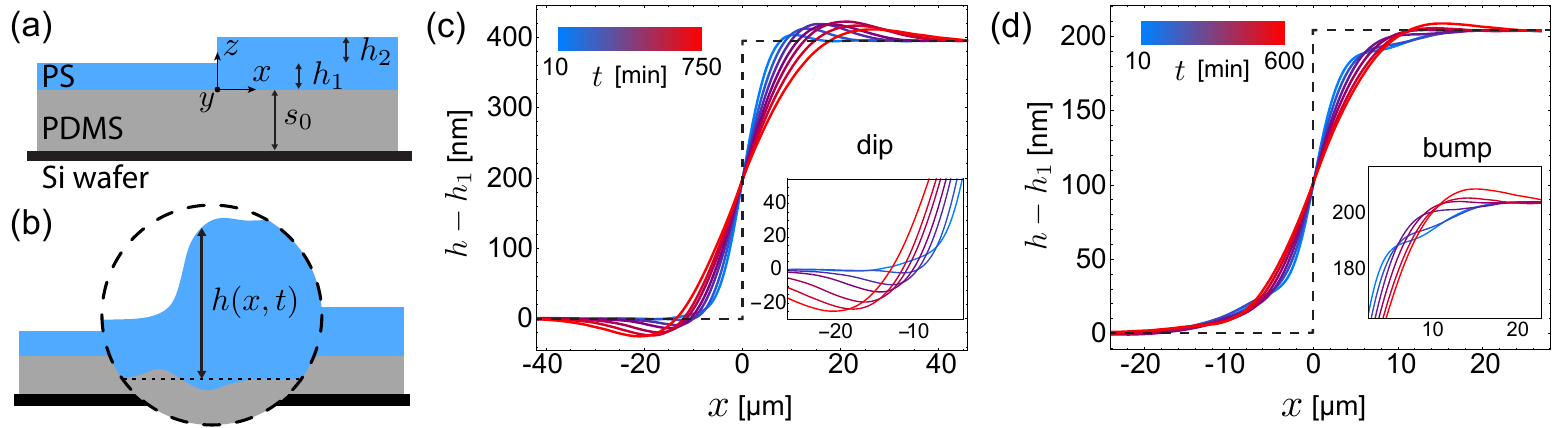}
\caption{\label{fig:schematics_and_profiles} (a) Schematics of the initial geometry: a stepped liquid polystyrene (PS) film is supported by an elastic layer of polydimethylsiloxane (PDMS). (b) Schematics of the levelling dynamics: the liquid height $h$ depends on the horizontal position $x$ and the time $t$. The elastic layer deforms due to the interaction with the liquid. (c) Experimental profiles of the liquid-air interface during levelling at $T_{\textrm{a}} = 140\,^\circ$C on 10:1 PDMS. The initial step has $h_1 = h_2 = 395\,$nm. The inset shows a close-up of the dip region. (d) Experimental profiles during levelling at $T_{\textrm{a}} = 140\,^\circ$C on the softer 40:1 PDMS. The initial step has $h_1 = h_2 = 200\,$nm. The inset shows a magnification of the bump region. Dashed lines in (c) and (d) indicate the initial condition.}
\end{figure*}
Interactions of solids and fluids are often pictured by the flapping of a flag in the wind, the oscillating motion of an open hosepipe or that of a fish fin in water, a set of examples in which the inertia of the fluid plays an essential role.
In contrast, at small scales, and more generally for low-Reynolds-number (Re) flows, fluid-solid interactions involve viscous forces rather than inertia. 
Of particular interest are the configurations where a liquid flows along a soft wall, \textit{i.e.} an elastic layer that can deform under the action of pressure and viscous stresses. For instance, when a solid object moves in a viscous liquid close to an elastic wall, the intrinsic symmetry of the Stokes equations that govern low-Re flows breaks down.
This gives rise to a qualitatively different -- elastohydrodynamical -- behaviour of the system in which the moving object may experience lift or oscillating motion \citep{skotheim2004,salez2015elastohydrodynamics,saintyves2016}, and a swimmer can produce a net thrust even by applying a time-reversible stroke \citep{trouilloud2008}, in apparent violation of the so-called scallop theorem \citep{purcell1977}. 
This coupling of viscous dynamics and elastic deformations is particularly significant in lubrication problems, such as the ageing of mammalian joints and their soft cartilaginous layers \citep{mccutchen1983lubrication}, or roll-coating processes involving rubber-covered rolls \citep{coyle1988roll-coating}, among others.

When adding a liquid-vapour interface, capillary forces may come into play, thus allowing for elastocapillary interactions. The latter have attracted a lot of interest in the past decade \citep{roman2010,Andreotti2016,style2016elastocapillarity}. In order to enhance the effect of capillary forces, the elastic object has to be either slender or soft. 
The first case, in which the elastic structure is mainly bent by surface tension, has been explored to explain and predict features like deformation and folding of plates, wrapping of plates (capillary origami) or fibers around droplets, and liquid imbibition between fibres \citep{py2007,antkowiak2011,duprat2012,Nadermann2013,Schulman:2015aa,Paulsen2015,Elettro2016,Schulman2017}. 
The second case involves rather thick substrates, where capillary forces are opposed by bulk elasticity. A common example is that of a small droplet sitting on a soft solid. 
\citet{lester1961} has been the first to recognize that the three-phase contact line can deform the substrate by creating a ridge. 
Despite the apparent simplicity of this configuration, the substrate deformation close to the contact line represents a challenging problem because of the violation of the classical Young's construction for the contact angle, the singularity of the displacement field at the contact line, and the difficulty to predict the exact shape of the capillary ridge. 
In the last few years, several theoretical and experimental works have contributed to a better fundamental understanding of this static problem \citep{pericet-camara2008, jerison2011, marchand2012contact, limat2012, style2013, lubbers2014}, recently extended by the dynamical case of droplets moving along a soft substrate \citep{Style2013durotaxis, karpitschka2015droplets, Karpitschka:2016aa}. 

Besides, another class of problems -- the capillary levelling of thin liquid films on rigid substrates, or in freestanding configurations -- has been studied in the last few years using thin polymer films featuring different initial profiles, such as steps, trenches, and holes \citep{mcgraw2011, Teisseire2011, mcgraw2012, baumchen2013, backholm2014capillary,Ilton2016}. 
From the experimental point of view, this has been proven to be a reliable system due to systematic reproducibility of the results and the possibility to extract rheological properties of the liquid \citep{rognin2011viscosity, mcgraw2013}. 
A theoretical framework, based on Stokes flow and the lubrication approximation, results in the so-called thin-film equation \citep{oron1997}, which describes the temporal evolution of the thickness profile. 
From this model, characteristic self-similarities of the levelling profiles, as well as numerical \citep{Salez:2012aa} and analytical \citep{salez2012,Benzaquen2013} solutions have been derived, which were found in excellent agreement with the experimental results. Furthermore, coarse-grained molecular dynamics models allowed to extend the framework of capillary levelling by offering local dynamical insights and probing viscoelasticity \citep{Tanis2017}.

In this article, by combining the two classes of problems above -- elastocapillarity and capillary levelling -- we design a novel dynamical elastocapillary situation free of any three-phase contact line. Specifically, we consider a setting in which a thin layer of viscous liquid with a non-flat thickness profile is supported onto a soft foundation. The liquid-air interface has a spatially varying curvature that leads to gradients in Laplace pressure, which drive flow coupled to substrate deformation. The resulting elastocapillary levelling might have practical implications in biological settings and nanotechnology.

\section{Experimental setup}
First, polydimethylsiloxane (PDMS, Sylgard 184, Dow Corning) is mixed with its curing agent in ratios varying from 10:1 to 40:1. In order to decrease its viscosity, liquid PDMS is diluted in toluene (Sigma-Aldrich, Chromasolv, purity > 99.9\%) to obtain a 1:1 solution in weight. 
The solution is then poured on a $15 \times 15$ mm Si wafer (Si-Mat, Germany) and spin-coated for 45\,s at 12.000\,RPM. 
The sample is then immediately transferred to an oven and kept at $75\,^\circ$C for 2 hours. 
The resulting elastic layer has a thickness $s_0 = 1.5 \pm 0.2 \; \mu$m, as obtained from atomic force microscopy (AFM, Multimode, Bruker) data. The Young's modulus of PDMS strongly depends on the ratio of base to cross-linker, with typical values of $E = 1.7 \pm 0.2\,$MPa for 10:1 ratio, $E = 600 \pm 100\,$kPa for 20:1 and $E = 50 \pm 20\,$kPa for 40:1 \citep{brown2005evaluation, hemmerle2016}.

In order to prepare polystyrene (PS) films exhibiting non-constant curvatures, we employ a technique similar to that described in \citep{mcgraw2011}. Solutions of 34\,kg/mol PS (PSS, Germany, polydispersity < 1.05) in toluene with typical concentrations varying between 2\% and 6\% are made. 
A solution is then spin-cast on a freshly cleaved mica sheet (Ted Pella, USA) for about 10\,s, with typical spinning velocities on the order of a few thousands RPM. After the rapid evaporation of the solvent during the spin-coating process, a thin (glassy) film of PS is obtained, with a typical thickness of $200 - 400$\,nm.

To create the geometry required for the levelling experiment, a first PS film is floated onto a bath of ultra-pure (MilliQ) water. 
Due to the relatively low molecular weight of the PS employed here, the glassy film spontaneously ruptures into several pieces. 
A second (uniform) PS film on mica is approached to the surface of water, put into contact with the floating PS pieces and rapidly released as soon as the mica touches the water. 
That way a collection of PS pieces is transferred onto the second PS film, forming a discontinuous double layer that is then floated again onto a clean water surface. 
At this stage, a sample with the elastic layer of PDMS is put into the water and gently approached to the floating PS from underneath. 
As soon as contact between the PS film and the PDMS substrate is established, the sample is slowly released from the bath. 
Finally, the initial configuration depicted in Fig.~\ref{fig:schematics_and_profiles}(a) is obtained.
For a direct comparison with capillary levelling on rigid substrates, we also prepared stepped PS films of the same molecular weight on freshly cleaned Si wafers (Si-Mat, Germany) using the same transfer procedure.

Using an optical microscope we identify spots where isolated pieces of PS on the uniform PS layer display a clean and straight interfacial front. A vertical cross-section of these spots corresponds to a stepped PS-air interface, which is invariant in the $y$ dimension (see Fig.~\ref{fig:schematics_and_profiles}(a) for a sketch of this geometry). 
Using AFM, the 3D shape of the interface is scanned and a 2D profile is obtained by averaging along $y$. From this profile the initial height of the step $h_2$ is measured. The sample is then annealed at an elevated temperature $T_{\textrm{a}} = 120-160\,^\circ$C (above the glass-transition temperature of PS) using a high-precision heating stage (Linkam, UK). During this annealing period the liquid PS flows. Note that on the experimental time scales and for the typical flow velocities studied here the PS is well described by a Newtonian viscous fluid~\citep{mcgraw2011,mcgraw2012, baumchen2013,backholm2014capillary,Ilton2016} (viscoelastic and non-Newtonian effects are absent since the Weissenberg number $\textrm{Wi} \ll 1$ and the Deborah number $\textrm{De} \ll 1$). After a given annealing time $t$, the sample is removed from the heating stage and quenched at room temperature (below the glass-transition temperature of PS). The 3D PS-air interface in the zone of interest is scanned with the AFM and a 2D profile is again obtained by averaging along $y$. This procedure is repeated several times in order to monitor the temporal evolution of the height $h(x,t)$ of the PS-air interface (defined with respect to the undeformed elastic-liquid interface, see Fig.~\ref{fig:schematics_and_profiles}(b)). 
At the end of each experiment, the thickness $h_1$ of the uniform PS layer is measured by AFM.  

\section{Experimental results and discussion}
\subsection{Profile evolution}
The temporal evolutions of two typical profiles are reported in Figs.~\ref{fig:schematics_and_profiles}(c),(d), corresponding to films that are supported by elastic foundations made of 10:1 PDMS and 40:1 PDMS, respectively. 
As expected, the levelling process manifests itself in a broadening of the initial step over time. 
In all profiles, three main regions can be identified (from left to right): a region with positive curvature (negative Laplace pressure in the liquid), an almost linear region around $x = 0$ (zero Laplace pressure) and a region of negative curvature (positive Laplace pressure in the liquid). 
These regions are surrounded by two unperturbed flat interfaces exhibiting $h = h_1$ and $h = h_1 + h_2$. 
In analogy with earlier works on rigid substrates \citep{mcgraw2012}, we refer to the positive-curvature region of the profile as the \textit{dip}, and the negative-curvature region as the \textit{bump}. Close-up views of those are given in the insets of Figs.~\ref{fig:schematics_and_profiles}(c),(d).  

The decrease of the slope of the linear region is a direct consequence of levelling.  
A less intuitive evolution is observed in the bump and dip regions. For instance, in the first profile of Fig.~\ref{fig:schematics_and_profiles}(c), recorded after 10\,min of annealing, a bump has already emerged while a signature of a dip cannot be identified yet. As the interface evolves in time, a dip appears and both the bump and the dip grow substantially. At a later stage of the evolution, the height of the bump and the depth of the dip eventually saturate.
This vertical evolution of the bump and the dip is at variance with what has been observed in the rigid-substrate case \citep{mcgraw2011,mcgraw2012}, where the values of the maximum and the minimum are purely dictated by $h_1$ and $h_2$ and stay fixed during the experimentally accessible evolution. 
That specific signature of the soft foundation is even amplified for PS levelling on the softer (40:1 PDMS) foundation, see Fig.~\ref{fig:schematics_and_profiles}(d). The evolution of the bump and dip results from the interaction between the liquid and the soft foundation. Indeed, the curvature gradients of the liquid-air interface give rise to Laplace pressure gradients that drive the flow. 
The pressure and flow fields both induce elastic deformations in the substrate. Intuitively, the negative Laplace pressure below the dip results in a traction that pulls upwards on the PDMS substrate, while the positive Laplace pressure below the bump induces a displacement in the opposite direction. In addition, a no-slip condition at the solid-liquid interface coupled to the flow induces an horizontal displacement field in the PDMS substrate. These displacements of the foundation act back on the liquid-air interface by volume conservation. 
According to this picture, the displacement of the solid-liquid interface is expected to tend to zero over time, since the curvature gradients of the liquid-air interface and the associated flow decrease.
\begin{figure}
\includegraphics[width=\columnwidth]{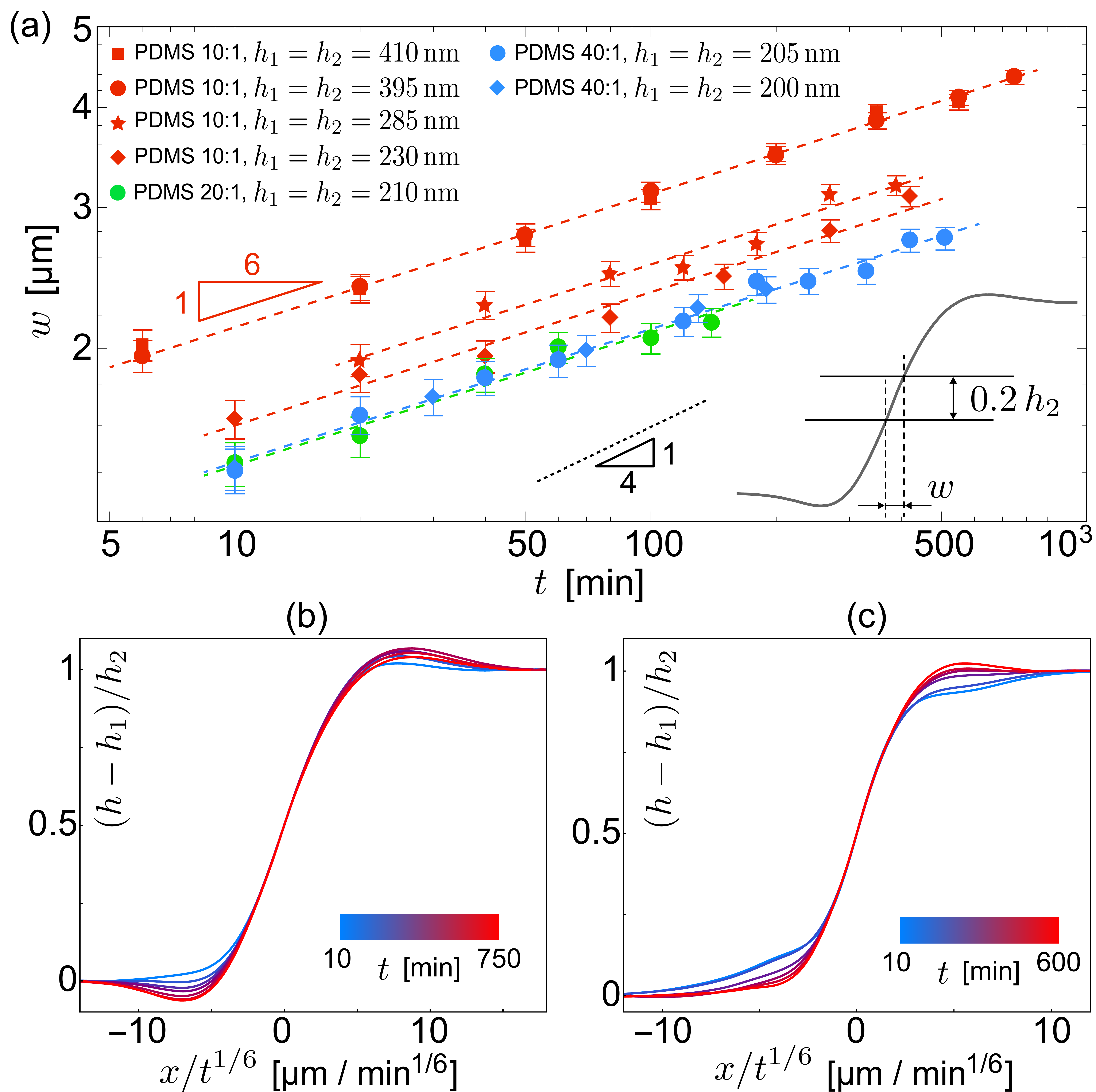}
\caption{\label{fig:width_and_profilesSelf} (a) Experimental evolution of the profile width $w$ (proportional to the lateral extent of the linear region as displayed in the inset) as a function of time $t$, in log-log scale, for samples involving different liquid-film thicknesses and substrate elasticities. All datasets seem to exhibit a $t^{1/6}$ power law. The slope corresponding to a $t^{1/4}$ evolution (rigid-substrate case) is displayed for comparison. (b) Experimental levelling profiles on 10:1 PDMS from Fig.~\ref{fig:schematics_and_profiles}(c) with the horizontal axis rescaled by $t^{-1/6}$. (c) Same rescaling applied for the levelling profiles on 40:1 PDMS shown in Fig.~\ref{fig:schematics_and_profiles}(d).}
\end{figure}

\subsection{Temporal evolution of the profile width}
The capillary levelling on a rigid substrate possesses an exact self-similar behaviour in the variable $x / t ^{1/4}$, leading to a perfect collapse of the rescaled height profiles of a given evolution \citep{mcgraw2012}. 
In contrast, for a soft foundation, no collapse of the profiles is observed (not shown) when the horizontal axis $x$ is divided by $t^{1/4}$. 

To determine whether another self-similarity exists or not, we first quantify the horizontal evolution of the profile by introducing a definition of its width (see Fig.~\ref{fig:width_and_profilesSelf}(a), inset): $w(t) = x (h = h_1 + 0.6 \, h_2) - x (h = h_1 + 0.4 \, h_2)$. With this definition, only the linear region of the profile matters and the peculiar shapes of the dip and bump do not affect the value of $w$.
The temporal evolution of $w$ was measured in several experiments, featuring different values of $h_1$, $h_2$ as well as three stiffnesses of the soft foundation. First, the absolute value of $w$ at a given time is larger for thicker liquid films, as expected since more liquid can flow. 
Secondly, the data plotted in Fig.~\ref{fig:width_and_profilesSelf}(a) clearly shows that in all these experiments the width increases as $w \sim t^{1/6}$. Equivalently, dividing the horizontal axis $x$ by $t^{1/6}$ leads to a collapse of all the linear regions of the profiles, as shown in Figs.~\ref{fig:width_and_profilesSelf}(b),(c). However, while allowing for the appreciation of the vertical evolution of the bump and dip, the non-collapse of the full profiles indicates the absence of true self-similarity in the problem. Nevertheless, we retain that for practical purposes associated with elastocapillary levelling, the $w \sim t^{1/6}$ scaling encompasses most of the evolution in terms of flowing material. 

\subsection{The role of viscosity}
\begin{figure}
\includegraphics[width=\columnwidth]{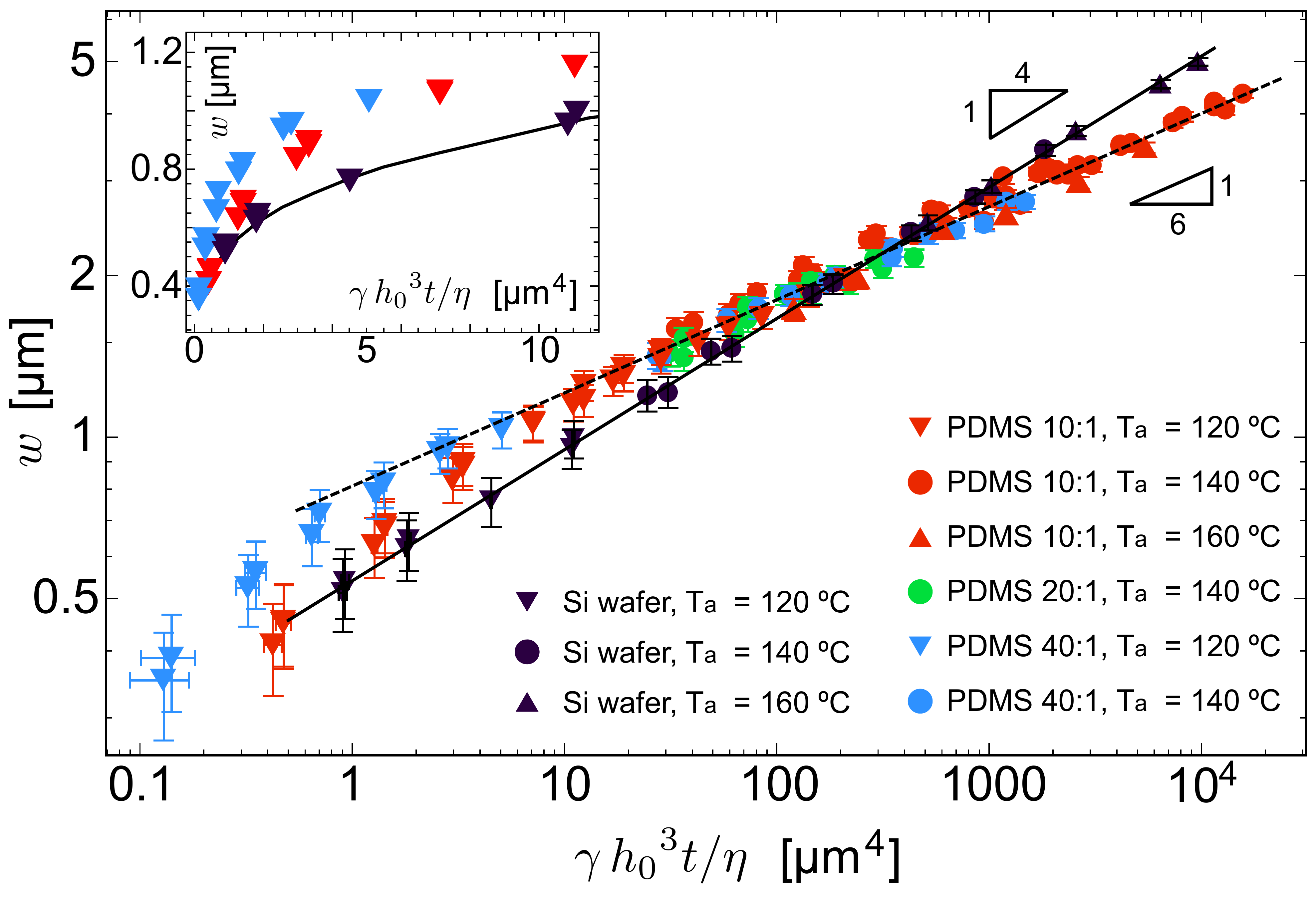}
\caption{\label{fig:width_rescaled} Experimental profile width $w$ (see Fig.~\ref{fig:width_and_profilesSelf}(a), inset) as a function of $\gamma {h_0}^3 t / \eta$ (see definitions in text), in log-log scale, for all the different samples and temperatures. Experiments for 10:1 (red), 20:1 (green) and 40:1 (blue) PDMS substrates, as well as annealing temperatures $T_{\textrm{a}} = 120 ^\circ$C (down triangle), $140 ^\circ$C (circle), $160 ^\circ$C (up triangle) are displayed. Most of the data collapses on a single curve of slope $1/6$ (dashed line). The data for capillary levelling on rigid substrates (black symbols) are shown for comparison and collapse on a single curve of slope $1/4$ (solid line). The inset displays a close-up of the early-time regime in linear representation.}
\end{figure}
The impact of the soft foundation on the levelling dynamics depends on two essential aspects: the stiffness of the foundation and how strongly the liquid acts on it. 
The first aspect is constant, and controlled by both the Young's modulus $E$ and the thickness $s_0$ of the (incompressible) PDMS layer, the former being fixed by the base-to-cross-linker ratio. 
The second aspect is ultimately controlled by the Laplace pressure, which is directly related to the curvature of the liquid-air interface. Even for a single experiment, the amplitude of the curvature field associated with the profile evolves along time, from large values at early times, to small ones at long times when the profile becomes almost flat. Thus, we expect the relative impact of the soft foundation to change over time. 

This time dependence can be explored by adjusting the PS viscosity. Indeed, the latter strongly decreases for increasing annealing temperature, while the other quantities remain mostly unaffected by this change. 
Hence, the levelling dynamics can be slowed down by performing experiments at lower annealing temperature, in order to investigate the dynamics close to the initial condition, and accelerated at higher annealing temperature in order to access the late-stage dynamics. 
Here, we report on experiments at $120\,^\circ \mathrm{C}$ (high viscosity) and  $160\,^\circ \mathrm{C}$ (low viscosity) and compare the results to our previous experiments at $140\,^\circ \mathrm{C}$. 

Following lubrication theory \citep{oron1997}, the typical time scale of a levelling experiment is directly fixed by the capillary velocity $\gamma/\eta$, where $\gamma$ denotes the PS-air surface tension and $\eta$ the PS viscosity, as well as the thickness $h_0=h_1+h_2/2$ of the PS film.
In Fig.~\ref{fig:width_rescaled}, the experimental profile width is plotted as a function of $\gamma {h_0}^3 t / \eta$  \citep{mcgraw2012}, for experiments involving different liquid film thicknesses, substrate elasticities and annealing temperatures. Samples with PS stepped films on bare (rigid) Si wafers were used to measure the capillary velocity $\gamma / \eta$ at different annealing temperatures \citep{mcgraw2013}. 
In these calibration measurements, the profile width follows a $t^{1/4}$ power law, as expected \citep{mcgraw2012}. 
In contrast, for the experiments on elastic foundations, two different regimes might be distinguished: 
for  $\gamma {h_0}^3 t / \eta$ larger than $\sim5\ \mu$m$^4$, the width follows a $t^{1/6}$ power law and all datasets collapse onto a single master curve over 3 to 4 orders of magnitude on the horizontal scale; for values of $\gamma {h_0}^3 t / \eta$ smaller than $\sim5\ \mu$m$^4$, the evolution depends on the elastic modulus and it appears that the softer the foundation the faster the evolution (see inset of Fig.~\ref{fig:width_rescaled}). 

\subsection{Vertical evolution of the dip and bump}
\begin{figure}
\includegraphics[width=.97\columnwidth]{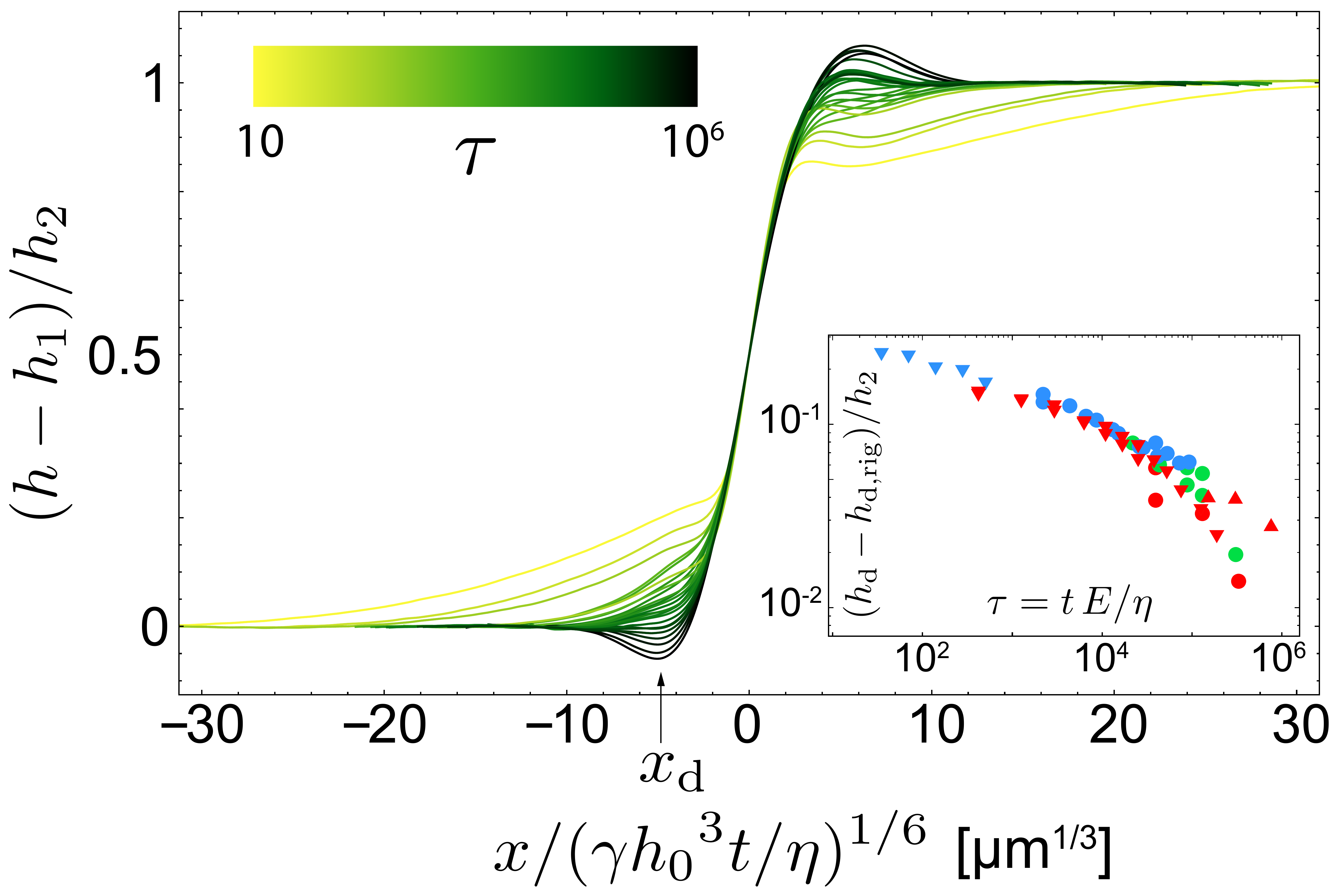}
\caption{\label{fig:vertical_evolution} Rescaled experimental profiles for all data displayed in Fig.~\ref{fig:width_rescaled}, colour-coded according to the dimensionless time $\tau = t E / \eta$. Inset: Evolution with $\tau$ of the normalized distance between the height $h_\mathrm{d}$ of the liquid-air interface at the dip position $x_\mathrm{d}$ and the corresponding value $h_\mathrm{d, rig}$ for the rigid case. Note that in all the experiments $h_1 = h_2$. Symbols are chosen to be consistent with Fig.~\ref{fig:width_rescaled}. }
\end{figure}
Guided by the previous discussion, we now divide the horizontal axis $x$ of all the height profiles in different experiments by the quantity $(\gamma {h_0}^3 t / \eta)^{1/6}$. 
As shown in Fig.~\ref{fig:vertical_evolution}, this rescaling leads to a collapse in the linear region of the profiles, while the dip and the bump regions display significant deviations from a universal collapse.

In order to characterize these deviations, we introduce the Maxwell-like viscoelastic time $\eta/E$ and define the dimensionless time $ \tau = E t / \eta$. 
This dimensionless parameter quantifies the role of the deformable substrate: experiments on softer foundations (lower $E$) or evolving slower (larger $\eta$) correspond to smaller values of $\tau$, and are therefore expected to show more pronounced elastic behaviours. 
As seen in Fig.~\ref{fig:vertical_evolution}, we find a systematic trend when plotting the experimental levelling profiles using the parameter $\tau$. Profiles with large $\tau$ (dark green and black) display clear bumps and dips, comparable in their vertical extents to the corresponding features observed on rigid substrates (not shown). In contrast, profiles with small $\tau$ (yellow and bright green) feature large deviations with respect to this limit.

The previous observation can be quantified by tracking the temporal evolution of the height of the liquid-air interface $h_\mathrm{d}(t) = h(x_\mathrm{d}, t)$ at the dip position $x_\mathrm{d}$, which we define as the (time-independent)  position at which the global minimum is located at the latest time of the levelling dynamics (see arrow in Fig.~\ref{fig:vertical_evolution}). The inset of Fig.~\ref{fig:vertical_evolution} displays the normalized difference between $h_\mathrm{d}$ and the corresponding value for a rigid substrate $h_\mathrm{d, rig}$, plotted as a function of $\tau$. 
We find that the parameter $\tau$ allows for a reasonable rescaling of the data. 
As anticipated, the difference between levelling on rigid and soft substrates decreases monotonically as a function of this dimensionless time. 
For small $\tau$, the difference can be larger than $20\%$ of the liquid film thickness, while for large $\tau$ it drops to less than $1\%$, which corresponds to the vertical resolution of the AFM.

\section{Theoretical modelling}
\subsection{Model and solutions}
\begin{figure}
\includegraphics[width=.94\columnwidth]{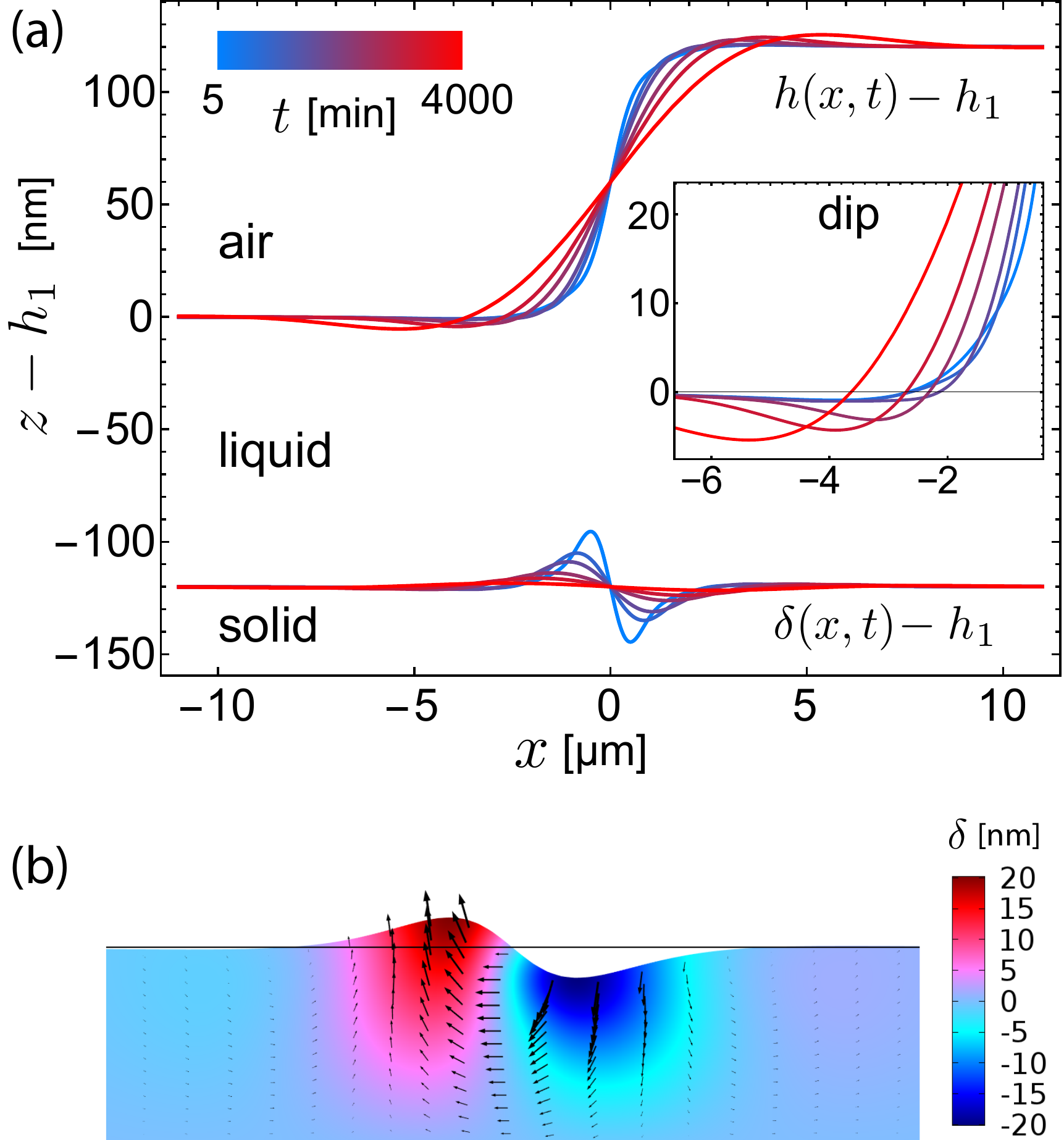}
\caption{\label{fig:model_profiles} (a) Theoretical profiles for the liquid-air interface $z=h(x,t)$ and the solid-liquid interface $z=\delta(x,t)$, both shifted vertically by $-h_1$. Here, we employ $s_0 = 2 \, \mu$m, $h_1 =h_2=2h_0/3=120$\,nm, $\mu = 25$\,kPa, $\gamma= 30\,$mN/m, $\eta = 2.5 \times 10^{6}$\,Pa\,s. The inset displays a close-up of the dip region. (b) Finite-element simulation (COMSOL) of the solid's total displacement (black arrows) and its vertical component $\delta$ (color code). The result has been obtained by imposing the Laplace pressure field corresponding to the first profile in (a) to a slab of elastic material exhibiting comparable geometrical and mechanical properties as in (a). The maximal displacement of $22$\,nm is in good agreement with the theoretical prediction shown in (a).}
\end{figure}
We consider an incompressible elastic slab atop which a viscous liquid film with an initial stepped liquid-air interface profile is placed. The following hypotheses are retained: i) the height $h_2$ of the step is small as compared to the thickness $h_0=h_1+h_2/2$ of the (flat) equilibrium liquid profile; ii) the slopes at the liquid-air interface are small, such that the curvature of the interface can be approximated by $\partial_x^{\,2}h$; iii) the lubrication approximation applies in the liquid, \textit{i.e.}\ typical vertical length scales are much smaller than horizontal ones; iv) the components of the displacement field in the elastic material are small as compared to the thickness of the elastic layer (linear elastic behaviour); v) the elastic layer is incompressible (valid assumption for PDMS). Note that the hypotheses i) to iii) have been successfully applied in previous work on the levelling dynamics of a stepped perturbation of a liquid film placed on a rigid substrate \citep{salez2012}.

Below, we summarize the model, the complete details of which are provided in the Supplementary Material. The main difference with previous work \citep{salez2012} is the coupling of fluid flow and pressure to elastic deformations of the substrate. The Laplace pressure is transmitted by the fluid and gives rise to a vertical displacement $\delta(x,t)$ of the solid-liquid interface, and thus an horizontal displacement $u_{\textrm{s}}(x,t)$ of the latter by incompressibility. Consequently, the no-slip condition at the solid-liquid interface implies that a fluid element in contact with the elastic surface will have a non-zero horizontal velocity $\partial_tu_{\textrm{s}}$. In addition, we assume no shear at the liquid-air interface.
After linearization, the modified thin-film equation reads:
\begin{equation}\label{TFE}
\frac{\partial{\Delta}}{\partial{t}}+\frac{\partial{}}{\partial{x}} 
\left[ -\frac{h_0^{\,3}}{3\eta}\frac{\partial{p}}{\partial{x}}+h_0\frac{\partial{u_{\textrm{s}}}}{\partial{t}} \right] = 0\ ,
\end{equation}
where $\Delta(x,t)=h(x,t)-\delta(x,t)-h_0$ is the excess thickness of the liquid layer with respect to the equilibrium value $h_0$.
The excess pressure $p(x,t)$ in the film, with respect to the atmospheric value, is given by the (small-slope) Laplace pressure:
\begin{equation}\label{Laplace}
p\simeq-\gamma \frac{\partial^2{(\Delta+\delta)}}{\partial{x^2}}\ .
\end{equation}
Furthermore, the surface elastic displacements are related to the pressure field through:
\begin{equation}\label{Greens}
\delta=-\frac{1}{\sqrt{2\pi}\mu}\int_{-\infty}^{\infty}k(x-x^{\prime})p(x^{\prime},t)\,\textrm{d}x',
\end{equation}
\begin{equation}\label{Greens2}
u_{\textrm{s}}=-\frac{1}{\sqrt{2\pi}\mu}\int_{-\infty}^{\infty}k_{\textrm{s}}(x-x^{\prime})p(x^{\prime},t)\,\textrm{d}x',
\end{equation}
where $\mu=E/3$ is the shear modulus of the incompressible substrate, and where $k(x)$ and $k_{\textrm{s}}(x)$ are the Green's functions (see Supplementary Material) for the vertical and horizontal surface displacements, \textit{i.e.}\ the fundamental responses due to a line-like pressure source of magnitude $-\sqrt{2\pi}\mu$  acting on the surface of the infinitely long elastic layer.

Equations~(\ref{TFE})-(\ref{Greens2}) can be solved analytically using Fourier transforms (see Supplementary Material), and we obtain:
\begin{equation}\label{FT}
\begin{split}
\tilde{\Delta}(\lambda,t)=-\frac{h_2}{2i\lambda}\sqrt{\frac{2}{\pi}}\\
\exp \left[-\left(\frac{\gamma\lambda^4h_0^{\,3}}{3\eta}\right)\frac{t}{1+(\gamma\lambda^2/\mu)(\tilde{k}+i\lambda h_0\tilde{k_{\textrm{s}}})} \right]\ ,
\end{split}
\end{equation}
\begin{equation}\label{FT2}
\tilde{\delta}=\frac{-\tilde{k}}{\mu}\frac{\gamma\lambda^2\tilde{\Delta}}{\left[1+(\gamma\lambda^2/\mu)\tilde{k}\right ]}\ ,
\end{equation}
where $\tilde{}$ denotes the Fourier transform of a function and $\lambda$ is the conjugated Fourier variable, \textit{i.e.}\ $\tilde{f}(\lambda)=\frac{1}{\sqrt{2\pi}}\int_{-\infty}^{\infty}f(x)e^{i\lambda x}\,\textrm{d}x$.
The vertical displacement $h(x,t)-h_0$ of the liquid-air interface with respect to its final state is then determined by summing the inverse Fourier transforms of Eqs.~(\ref{FT}) and (\ref{FT2}).

Figure~\ref{fig:model_profiles}(a) displays the theoretical profiles of both the liquid-air interface $z=h(x,t)$ and the solid-liquid interface $z=\delta(x,t)$, for a stepped liquid film with thicknesses $h_1 = h_2 =2h_0/3 =120$\,nm, supported by a substrate of stiffness $\mu = 25$\,kPa and thickness $s_0 = 2 \, \mu$m.
The viscosity $\eta = 2.5 \times 10^{6}$\,Pa\,s is adapted to the PS viscosity at the annealing temperature $T_{\textrm{a}} = 120 ^\circ$C in the experiment. 
The PS-air surface tension is fixed to $\gamma=30\,$mN/m \citep{brandrup1989polymer}. We find that the profiles predicted by this model reproduce some of the key features observed in our experiments. In particular, the evolutions of the bump and dip regions in the theoretical profiles (see Fig.~\ref{fig:model_profiles} inset) qualitatively capture the characteristic behaviours recorded in the experiment (see Fig.~\ref{fig:schematics_and_profiles}(c) inset). 

An advantage of this theoretical approach is the possibility to extract information about the deformation of the solid-liquid interface. 
As shown in Fig.~\ref{fig:model_profiles}, the substrate deforms mainly in the bump and dip regions, as a result of their large curvatures. 
The maximal vertical displacement of the solid-liquid interface in this example is $\sim25$\,nm, and it reduces over time, due to the levelling of the profile and the associated lower curvatures. 

\subsection{Evolution of the profile width}
\begin{figure}
\includegraphics[width=.99\columnwidth]{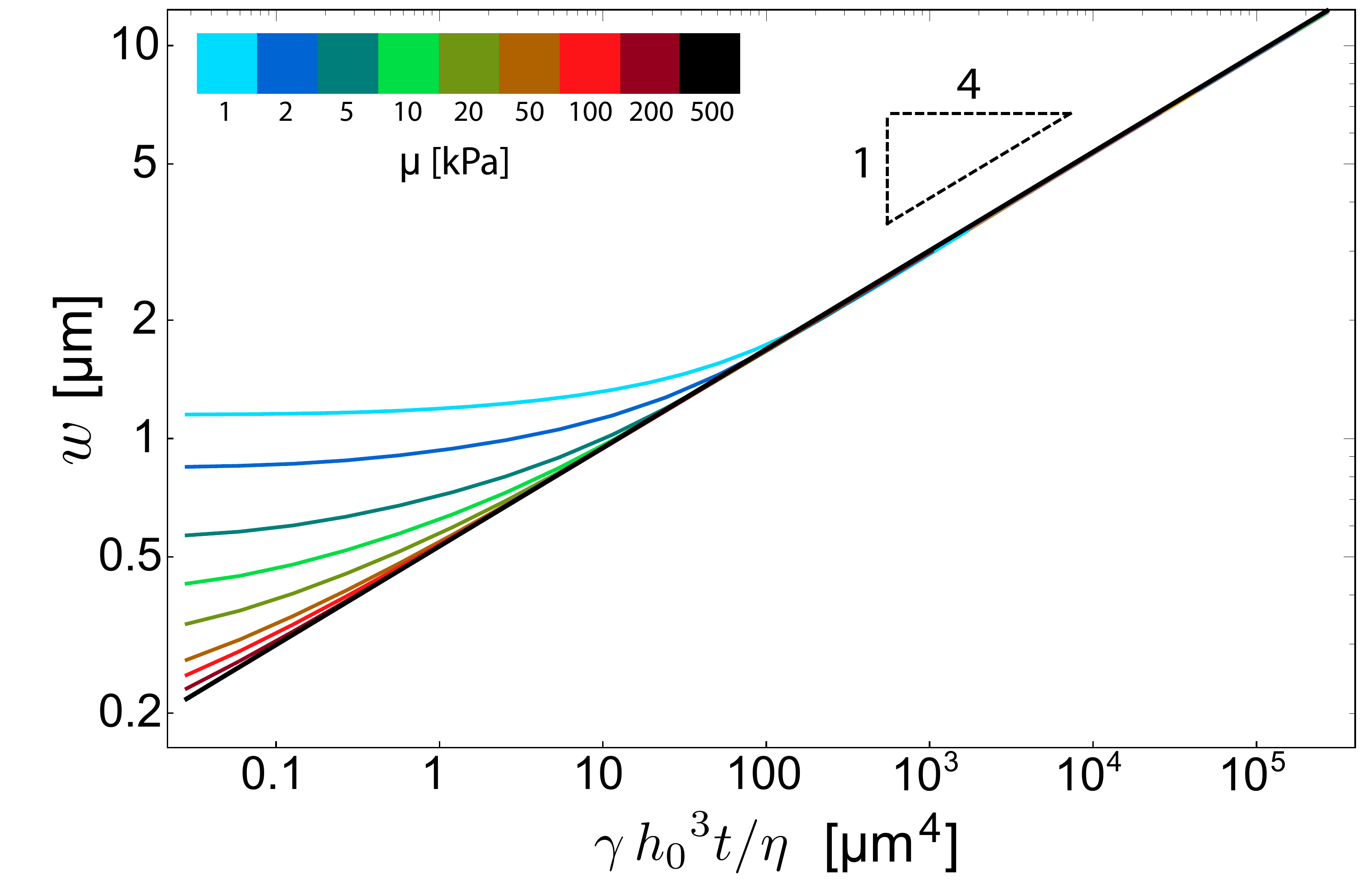}
\caption{\label{fig:model_width_evolution} Temporal evolution of the profile width (see definition in Fig.~\ref{fig:width_and_profilesSelf}(a), inset), in log-log scale, as predicted by the theoretical model, for different shear moduli, viscosities and liquid-film thicknesses. 
The $1/4$ power law corresponding to a rigid substrate is indicated. 
}
\end{figure}
The temporal evolution of the width $w$ (see Fig.~\ref{fig:width_and_profilesSelf}(a), inset) of the profiles was extracted from our theoretical model for a series of different parameters. Figure~\ref{fig:model_width_evolution} shows the theoretical width $w$ as a function of the quantity $\gamma {h_0}^3 t / \eta$ for all cases studied. With this rescaling, it is evident that the width of the theoretical profile depends strongly on elasticity at early times, while all datasets collapse onto a single curve at long times.
Moreover, this master curve exhibits a slope of $1/4$, and thus inherits a characteristic signature of capillary levelling on a rigid substrate.
The early-time data shows that the width is larger than on a rigid substrate, but with a slower evolution and thus a lower effective exponent. These observations are in qualitative agreement with our experimental data. 
However, interestingly, we do not recover in the experiments the predicted transition to a long-term rigid-like $1/4$ exponent, but instead keep a $1/6$ exponent (see Fig.~\ref{fig:width_rescaled}).

It thus appears that we do not achieve a full quantitative agreement between the theoretical and experimental profiles. 
The initially sharp stepped profile could possibly introduce an important limitation on the validity of the lubrication hypothesis. Indeed, while this is not a problem for the rigid case since the initial condition is rapidly forgotten \citep{Benzaquen2013}, it is not \textit{a priori} clear if and how elasticity affects this statement. We thus checked (see Supplementary Material) that replacing the lubrication approximation by the full Stokes equations for the liquid part does not change notably the theoretical results. We also checked that the linearization of the thin-film equation is not the origin of the aforementioned discrepancy: in a test experiment with $h_2\ll h_1$ on a soft substrate we observed the same characteristic features -- and especially the $1/6$ temporal exponent absent of the theoretical solutions -- as the ones reported for the $h_1\approx h_2$ geometry (see Supplementary Material). Besides, we note that while the vertical deformations of the elastic material (see Fig.~\ref{fig:model_profiles}) are small as compared to the thickness $s_0$ of the elastic layer in the experimentally accessible temporal range, the assumption of small deformations could be violated at earlier times without affecting the long-term behaviour at stake.

Finally, we propose a simplified argument to qualitatively explain the smaller transient exponent in Fig.~\ref{fig:model_width_evolution}. We assume that the vertical displacement $\delta(x,t)$ of the solid-liquid interface mostly translates the liquid above, such that the liquid-air interface displaces vertically by the same amount, following:  
\begin{equation}
\label{ash}
h (x,t) = h_{\textrm{r}}(x,t) + \delta (x,t) \;,
\end{equation}
where $h_{\textrm{r}}$ is the profile of the liquid-air interface that would be observed on a rigid substrate. 
Note that this simplified mechanism does not violate conservation of volume in the liquid layer. 
By deriving the previous equation with respect to $x$, and evaluating it at the center of the profile ($x=0$), we obtain an expression for the central slope of the interface: 
\begin{equation}
\partial_xh(0,t)  = \partial_xh_{\textrm{r}}(0,t) + \partial_x\delta(0,t) \;. 
\end{equation}
Due to the positive (negative) displacement of the solid-liquid interface in the region $x<0$ ($x > 0$), $\partial_x\delta(0,t)$ is always negative, as seen in Fig.~\ref{fig:model_profiles}. 
Therefore, we expect a reduced slope of the liquid-air interface in the linear region, which is in agreement with the increased width observed on soft substrates. 
Moreover, taking the second derivative of Eq.~(\ref{ash}) with respect to $x$ leads to:
\begin{equation}
\partial_{x}^{\,2}h (x,t) = \partial_{x}^{\,2}h_{\textrm{r}}(x,t) + \partial_{x}^{\,2}\delta(x,t) \;.
\end{equation}
In the dip region, $h_{\textrm{r}}(x,t)$ is convex in space (positive second derivative with respect to $x$), while $\delta(x,t)$ is assumed to be concave in space (negative second derivative with respect to $x$) up to some distance from the center (see Fig.~\ref{fig:model_profiles}). 
Therefore, the resulting curvature is expected to be reduced. A similar argument leads to the same conclusion in the bump region. This effect corresponds to a reduction of the Laplace pressure and, hence, of the driving force for the levelling process: the evolution is slower which translates into a smaller effective exponent.

\subsection{Finite-element simulations}
To check the validity of the predicted shape of the solid-liquid interface, we performed finite-element simulations using COMSOL Multiphysics. Starting from an experimental profile of the liquid-air interface at a given time $t$, the curvature and the resulting pressure field $p(x,t)$ were extracted. This pressure field was used as a top boundary condition for the stress in a 2D slab of an incompressible elastic material exhibiting a  comparable thickness and stiffness as in the corresponding experiment.
The slab size in the $x$ direction was chosen to be 20\,$\mu$m, which is large enough compared to the typical horizontal extent of the elastic deformation (see Fig.~\ref{fig:model_profiles}(a)). 
The bottom boundary of the slab was fixed (zero displacement), while the left and right boundaries were let free (zero stress). 
The deformation field predicted by these finite-element simulations is shown in Fig.~\ref{fig:model_profiles}(b) and found to be in quantitative agreement with our theoretical prediction. 

\section{Conclusion}
We report on the elastocapillary levelling of a thin viscous film flowing above a soft foundation. The experiments involve different liquid film thicknesses, viscosities, and substrate elasticities. 
We observe that the levelling dynamics on a soft substrate is qualitatively and quantitatively different with respect to that on a rigid substrate.
At the earliest times, the lateral evolution of the profiles is faster on soft substrates than on rigid ones, as a possible result of the ``instantaneous" substrate deformation caused by the capillary pressure in the liquid. 
Immediately after, this trend reverses: the lateral evolution of the profiles on soft substrates becomes slower than on rigid ones, which might be related to a reduction of the capillary driving force associated with the elastic deformation.
Interestingly, we find that the width of the liquid-air interface follows a $t^{1/6}$ power law over several orders of magnitude on the relevant scale, in sharp contrast with the classical $t^{1/4}$ law observed on rigid substrates. 

To the best of our knowledge, this system is a unique example of dynamical elastocapillarity that is not mediated by the presence of a contact line, but only by the Laplace pressure inside the liquid. Notwithstanding, this process is not trivial, since the coupled evolutions of both the liquid-air and solid-liquid interfaces lead to an intricate dynamics. 
Our theoretical approach, based on linear elasticity and lubrication approximation, is able to reproduce some observations, such as the typical shapes of the height profiles and the dynamics at short times. 

While some characteristic experimental features are captured by the model, a full quantitative agreement is still lacking to date. Given the careful validation of all the basic assumptions underlying our theoretical approach (\textit{i.e.} lubrication approximation, linearization of the thin-film equation, and linear elasticity), we hypothesise that additional effects are present in the materials/experiments. For instance, it remains unclear whether the physicochemical and rheological properties at the surface of PDMS films, which were prepared using conventional recipes, are correctly described by bulk-measured quantities~\citep{Andreotti2016}. We believe that further investigations of the elastocapillary levelling on soft foundations, using different elastic materials and preparation schemes, could significantly advance the understanding of such effects and dynamic elastocapillarity in general.

Finally, we would like to stress that the signatures of elasticity in the elastocapillary levelling dynamics are prominent even on substrates that are not very soft (bulk Young's moduli of the PDMS in the $\sim$\,MPa range) and for small Laplace pressures. 
In light of applications such as traction-force microscopy, where localised displacements of a soft surface are translated into the corresponding forces acting on the material, the elastocapillary levelling on soft substrates might be an ideal model system to quantitatively study surface deformations in soft materials with precisely-controlled pressure fields.

\section{Acknowledgments}
The authors acknowledge Stephan Herminghaus, Jacco Snoeijer, Anupam Pandey, Howard Stone, Martin Brinkmann, Corentin Mailliet, Pascal Damman, Kari Dalnoki-Veress, Anand Jagota, and Joshua McGraw for interesting discussions. The German Research Foundation (DFG) is acknowledged for financial support under grant BA~3406/2. V.B. acknowledges financial support from \'Ecole Normale Sup\'erieure. T.S. acknowledges financial support from the Global Station for Soft Matter, a project of Global Institution for Collaborative Research and Education at Hokkaido University. C.-Y. Hui acknowledges financial support from the U.S. Department of Energy, Office of Basic Energy Sciences, Division of Materials Sciences and Engineering under Award DE-FG02-07ER46463, and from the Michelin-ESPCI Paris Chair. O.B. acknowledges financial support from the Joliot ESPCI Paris Chair and the Total-ESPCI Paris Chair.
\bibliography{biblio}
\pagebreak
\widetext

\begin{center}
\textbf{\large Supplementary Material}
\end{center}
\setcounter{equation}{0}
\setcounter{figure}{0}
\setcounter{section}{0}
\setcounter{table}{0}
\setcounter{page}{1}
\makeatletter
\renewcommand{\thefigure}{S\@arabic\c@figure}
\renewcommand{\theequation}{S\@arabic\c@equation}
\renewcommand{\thesection}{S\@arabic\c@section}

\section{Introduction}
\begin{figure}[h!]
\includegraphics[scale=0.75]{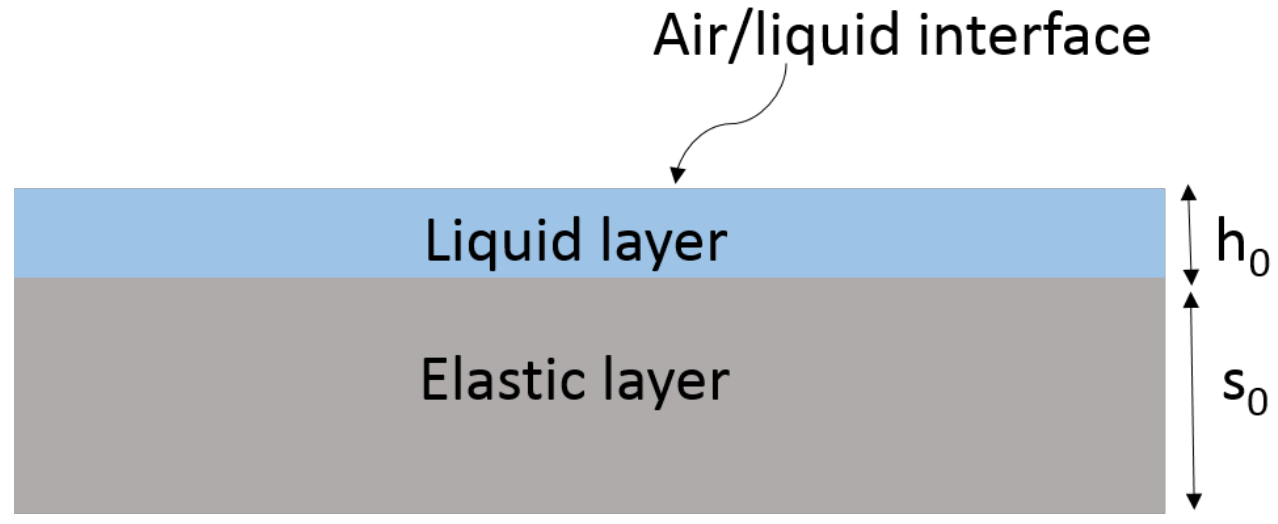}
\caption{Cross-sectional view of the reference equilibrium state, to which will be superimposed a deformation of the air-liquid interface at initial time ($t = 0$).}
\label{scheme}
\end{figure}

As a reference state, we consider a thin viscous film of height $h_0$ sitting on an incompressible elastic layer of thickness $s_0$ (see Fig.~\ref{scheme}). The elastic layer is itself placed atop a rigid substrate. 
We use a Cartesian coordinate system ($x$, $y$, $z$), with $z$ being the vertical coordinate. We assume the system to be infinite in the $x$ and $y$ directions. The surface tension of the air-liquid interface is denoted $\gamma$, the viscosity of the fluid (assumed to be Newtonian) $\eta$, and the shear modulus of the elastic material (assumed incompressible, \textit{i.e.} with a Poisson ratio of 1/2) $\mu$. At initial time, $t = 0$, we perturb the air-liquid interface by adding a step function $h(x) = h_2 H(x)$ with $H(x<0) = -1/2$ and $H(x>0) = 1/2$. We assume invariance in the $y$ direction, and that the step height $h_2$ is small compared with the reference height $h_0$. 

\section{Control experiment with a stepped perturbation}
\begin{figure}[h!]
\includegraphics[scale=0.5]{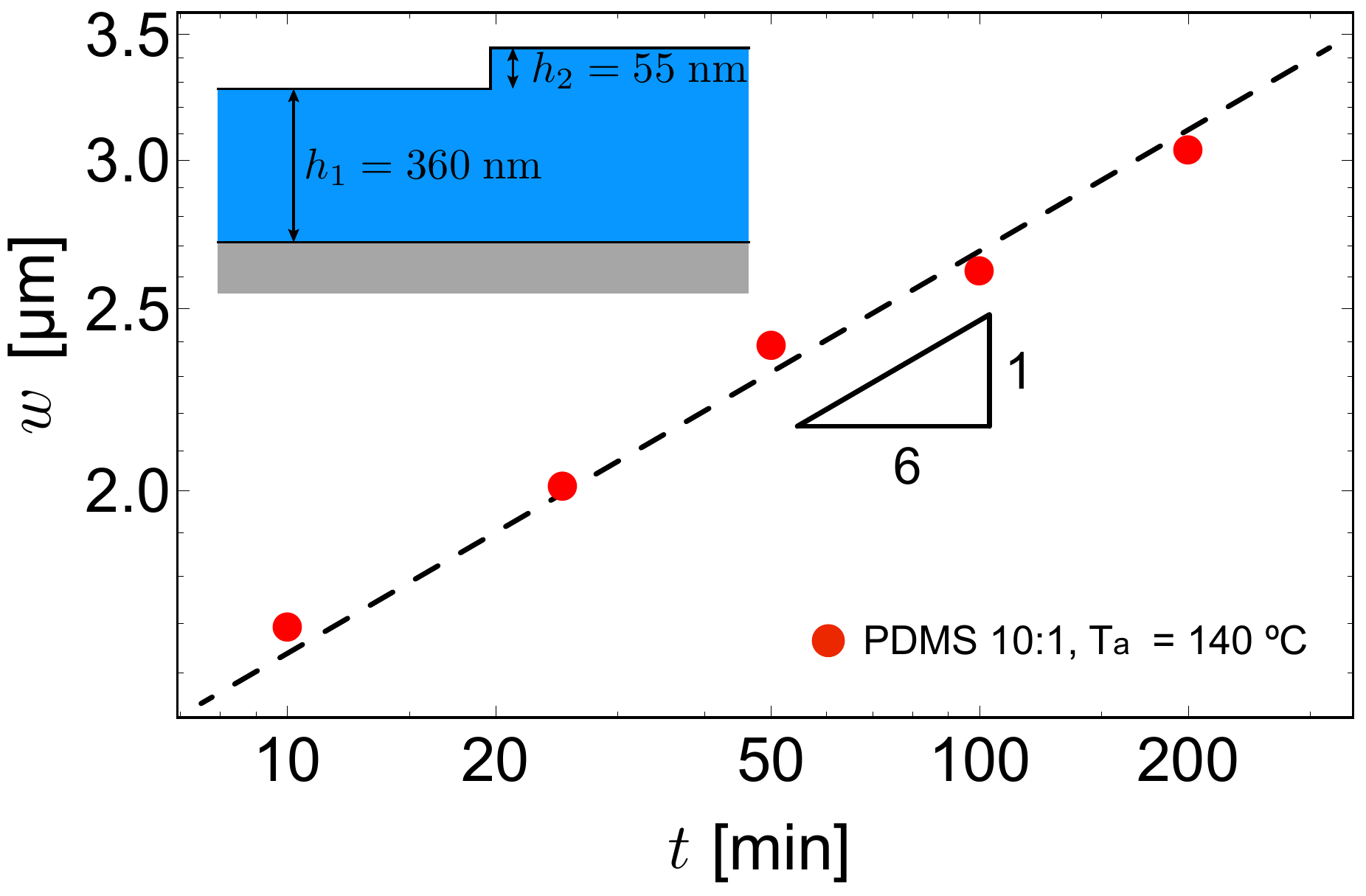}
\caption{Temporal evolution of the profile width $w$ (defined in the inset of Fig. 2(a)), in log-log scale, for an experiment with $h_2 \ll h_1$ (see inset). We used the same annealing temperature $T_{\textrm{a}}$ and PDMS substrate (in grey in the inset) as in the experiments reported in Figs. 1 and 2.}
\label{linear}
\end{figure}

The previous $h_2\ll h_0$ condition is not verified in our experiments (where $h_2 = h_1 = 2h_0/3$). However, we checked that this simplification in the model does not affect our general conclusions and is not the source of some discrepancy observed with the experiments. Indeed, in a test experiment with $h_2 \ll h_0$, we find that the profile width follows a $t^{1/6}$ power law with time $t$ (see Fig.~\ref{linear}), consistently with the $h_1 = h_2$ experimental case (see Figs. 2 and 3), and in contrast to the theoretical prediction (see Fig. 6).

\section{Lubrication-elastic Model}
\label{Lubrication}
\subsection{Lubrication description of the liquid layer}
As for the capillary levelling of a thin liquid film, of viscosity $\eta$, on a rigid substrate [39], we invoke the lubrication approximation which assumes that the typical horizontal length scale of the flow is much larger than the vertical one. As a result, at leading order, the vertical flow is neglected and the excess pressure field $p$ (with respect to the atmospheric pressure) does not depend on $z$. The incompressible Stokes' equations thus reduce to:
\begin{equation}
\frac{\partial p}{\partial x} = \eta \frac{\partial^2 v_x}{\partial z^2}\ , 
\end{equation}
which can be integrated in $z$ to get the horizontal velocity $v_x$. The main difference here with the previous model [39] is that the pressure acts on the elastic layer, giving rise to vertical and horizontal displacements of the liquid-elastic interface, $\delta (x,t)$ and $u_{\textrm{s}} (x,t)$ respectively. In addition, the no-slip condition at the liquid-elastic interface implies that a fluid particle in contact with the elastic surface will have a non-zero horizontal velocity $\partial u_{\textrm{s}}/\partial t$. Using this condition, the vanishing shear stress at the air-liquid interface, and invoking volume conservation, allow one to derive the following equation:
\begin{equation}
\frac{\partial \Delta}{\partial t} + \frac{\partial}{\partial x} \left[ - \frac{(h_0 + \Delta)^3}{3 \eta} \frac{\partial p }{\partial x} + (h_0 + \Delta) \frac{\partial u_{\textrm{s}}}{\partial t}   \right] = 0\ ,
\label{heigth_eq}
\end{equation}
where $\Delta (x,t) = h(x,t) - \delta(x,t) - h_0$ is the excess thickness of the liquid layer with respect to the equilibrium value $h_0$, and $h(x,t)$ is defined in Fig.~1(b). Since the pressure is independent of $z$, it is fixed by the proper boundary condition, \textit{i.e.} the Laplace pressure at the air-liquid interface (we neglect the non-linear term of the curvature at small slopes):
\begin{equation}
\label{Laplace}
p(x,t) = - \gamma \frac{\partial^2 h}{\partial x^2} = - \gamma \frac{\partial^2 (\Delta + \delta)}{\partial x^2}\ .
\end{equation}
Finally, as the perturbation is assumed to be small ($\Delta \ll h_0$), one can linearize Eq.~(\ref{heigth_eq}) and get the governing equation:
\begin{equation}
\frac{\partial \Delta}{\partial t} + \frac{\partial}{\partial x} \left[ - \frac{h_0^3}{3 \eta} \frac{\partial p }{\partial x} + h_0 \frac{\partial u_{\textrm{s}}}{\partial t}   \right] = 0\ .
\label{heigth_eq2}
\end{equation}

\subsection{Coupling with the elastic layer}
The surface displacements of the liquid-elastic interface are given by:
\begin{subequations}
\label{convolution}
\begin{equation}
\label{Conv1}
\delta(x,t) = - \frac{1}{ \sqrt{2 \pi} \mu} \int_{-\infty}^{\infty} \mathrm{d}x'\, k(x-x') p(x',t)\ ,
\end{equation}
\begin{equation}
\label{Conv2}
u_{\textrm{s}}(x,t) = - \frac{1}{ \sqrt{2 \pi} \mu} \int_{-\infty}^{\infty} \mathrm{d}x'\, k_{\textrm{s}}(x-x') p(x',t)\ ,
\end{equation}
\end{subequations}
where $k$ and $k_{\textrm{s}}$ are the Green's functions of the elastic problem (see Section~\ref{green}), corresponding to the vertical and horizontal displacements induced by a normal line load of magnitude $- \sqrt{2\pi }\mu$. We introduce the Fourier transform $\tilde f$ of a function $f$ with respect to its variable $x$ as: 
\begin{equation}
\tilde{f}(\lambda) = \frac{1}{\sqrt{2 \pi}} \int_{-\infty}^{\infty} \mathrm{d}x\, f(x) e^{i \lambda x}\ ,
\end{equation}
where $\lambda$ is the Fourier variable (\textit{i.e.} the angular wavenumber). Taking the Fourier transform of Eqs.~(\ref{Laplace}),~(\ref{heigth_eq2}), and~(\ref{convolution}), we obtain: 
\begin{equation}
\tilde{\delta} = - \frac{\tilde{p} \tilde{k}}{\mu} =\frac{-\tilde{k} \gamma \lambda^2}{\mu (1 + \tilde{k} \gamma \lambda^2/ \mu)}\tilde{\Delta}\ ,
\label{del}
\end{equation}
\begin{equation}
\tilde{u_{\textrm{s}}} =   - \frac{\tilde{p} \tilde{k_{\textrm{s}}}}{\mu} = \frac{-\tilde{k_{\textrm{s}}} \gamma \lambda^2}{\mu (1 + \tilde{k} \gamma \lambda^2 /\mu) } \tilde{\Delta}\ ,
\label{usdef}
\end{equation}
\begin{equation}
\frac{\partial \tilde{\Delta}}{\partial t} = - \Omega(\lambda) \tilde{\Delta}\ ,
\label{height_fou}
\end{equation}
and:
\begin{equation}
\Omega(\lambda) = \frac{\gamma \lambda^4 h_0^3}{3 \eta} \, \, \frac{1}{1 + (\gamma \lambda^2 / \mu) \left(\tilde{k} + i \lambda h_0 \tilde{k_{\textrm{s}}}   \right)}\ .
\end{equation}
The solution of Eq.~(\ref{height_fou}) is:
\begin{equation}
\tilde{\Delta}(\lambda,t) = \tilde{\Delta}(\lambda,0) \exp[-\Omega(\lambda) t] = -\frac{h_2}{2i\lambda} \sqrt{\frac{2}{\pi}} \, \exp[-\Omega(\lambda) t]\ ,
\end{equation}
where we have used the initial conditions $\Delta(x,0) = h_2 H(x)$ (see section I) and $\delta(x,0)=0$. Finally, using Eq.~(\ref{del}), one has: 
\begin{equation}
\label{Liquid_air}
\tilde{\Delta}(\lambda,t) + \tilde{\delta}(\lambda, t) = \frac{ \tilde{\Delta}(\lambda,t)}{1 + \tilde{k} \gamma \lambda^2 /\mu}\ .
\end{equation}
Therefore, once the Green's functions $\tilde{k}$ and $\tilde{k_{\textrm{s}}}$ are determined (see Section~\ref{green}), the displacement $h(x,t)-h_0=\Delta(x,t)+\delta(x,t)$ of the air-liquid interface with respect to its equilibrium position can be obtained by taking the inverse Fourier transform of Eq.~(\ref{Liquid_air}).

\subsection{Green's functions for the elastic layer}
\label{green}
We consider an incompressible and linear elastic layer of thickness $s_0$ supported on a rigid substrate (the latter is located at $z=-s_0$, see Fig.~1(a)). The deformation state of the elastic layer is that of plane strain, where the out-of-plane (\textit{i.e.} along $y$, see Fig.~1(a))) displacement is identically zero. The horizontal and vertical displacement fields, $u_x(x,z,t)$ and $u_z(x,z,t)$ respectively, are both fixed to zero at the rigid substrate:
\begin{subequations}
\begin{equation}
\label{BC_1}
u_x (x,-s_0,t) = 0\ ,
\end{equation}
\begin{equation}
\label{BC_2}
u_z (x,-s_0,t) = 0\ .
\end{equation}
\end{subequations}
On the other side of the layer, the liquid-elastic interface (located at $z=0$ at zeroth order in the perturbation, see Fig.~1(a)) is subjected to the lubrication pressure field $p(x,t)$, but we assume no shear which is valid at leading lubrication order. Therefore, one has:
\begin{subequations}
\begin{equation}
\label{BC_3}
\sigma_{zz}(x,0,t) = -p(x,t)\ ,
\end{equation}
\begin{equation}
\label{BC_4}
\sigma_{xz}(x,0,t) = 0\ .
\end{equation}
\end{subequations}
In plane strain, the stresses are given by the Airy stress function $\phi(x,z,t)$ which satisfies the spatial biharmonic equation. Specifically:
\begin{equation}
\sigma_{xx} = \frac{\partial^2 \phi}{\partial z^2} \quad \textrm{,} \quad \sigma_{zz} = \frac{\partial^2 \phi}{\partial x^2} \quad \textrm{and} \quad \sigma_{xz} = -\frac{\partial^2 \phi}{\partial x \partial z}\ .
\end{equation}
The generalized Hooke'€™s law for an incompressible material in plane strain reads:
\begin{subequations}
\begin{equation}
2 \mu \partial_z u_z = \sigma_{zz} - \Gamma\ ,
\end{equation}
\begin{equation}
2 \mu \partial_x u_x = \sigma_{xx} - \Gamma\ ,
\end{equation}
\begin{equation}
 \mu ( \partial_x u_z + \partial_z u_x)= \sigma_{xz}\ ,
\end{equation}
\end{subequations}
where $\Gamma(x,z,t)$ is the pressure needed to enforce incompressibility, that can be found using the incompressibility condition: 
\begin{equation}
\partial_x u_x + \partial_z u_z = 0 \quad  \Rightarrow \quad   \Gamma = \frac{\sigma_{xx} + \sigma_{zz}}{2}\ .
\end{equation}
Combining the above, and using the same Fourier-transform convention as in the previous section, we find the following relations:
\begin{equation}
\label{Stress_Airy}
\tilde{\sigma}_{xx} = \tilde{\phi}'' \quad \textrm{,} \quad \tilde{\sigma}_{zz} = -\lambda^2 \tilde{\phi} \quad \textrm{and} \quad \tilde{\sigma}_{xz} = i \lambda \tilde{\phi}'\ ,
\end{equation}
\begin{subequations}
\label{Disp_Airy}
\begin{equation}
2 \mu  \tilde{u}_z' = - \frac{\tilde{\phi}'' + \lambda^2 \tilde{\phi}}{2}\ ,
\end{equation}
\begin{equation}
- 2 i \lambda \mu  \tilde{u}_x =  \frac{\tilde{\phi}'' + \lambda^2 \tilde{\phi}}{2}\ ,
 \label{ux}
\end{equation}
\begin{equation}
 \mu ( -i\lambda \tilde{u}_z +\tilde{u}_x')= i \lambda \tilde{\phi}'\ ,
 \label{uz}
\end{equation}
\end{subequations}
where the prime denotes the partial derivative with respect to $z$. Taking the Fourier transform of the spatial biharmonic equation results in a fourth-order ordinary differential equation:
\begin{equation}
\lambda^4 \tilde{\phi} - 2\lambda^2 \tilde{\phi}'' + \tilde{\phi}'''' = 0\ ,
\end{equation}
whose general solution is:
\begin{equation}
\tilde{\phi}(\lambda,z,t)  = A(\lambda,t) \cosh(\lambda z) + B(\lambda,t) \sinh(\lambda z) + C(\lambda,t) z \cosh(\lambda z) + D(\lambda,t) z \sinh(\lambda z)\ .
\label{bihar_solution}
\end{equation}
The parameters $A$, $B$, $C$, $D$ are determined using the boundary conditions (Eqs.~(\ref{BC_1}), (\ref{BC_2}), (\ref{BC_3}), and (\ref{BC_4})) and the relations between the Airy stress function and the stresses/displacements (Eqs.~(\ref{Stress_Airy}) and (\ref{Disp_Airy})). After some algebra, we find:
\begin{equation}
A = \frac{\tilde{p}}{\lambda^2}   \quad \textrm{,} \quad B = \frac{\tilde{p}}{\lambda^2} \, \frac{\sinh(\lambda s_0)\cosh(\lambda s_0) - \lambda s_0}{\cosh^2(\lambda s_0) + (\lambda s_0)^2}  \quad \textrm{,} \quad C = - \lambda B  \quad \textrm{and} \quad D = -\frac{\tilde{p}}{\lambda} \, \frac{\cosh^2(\lambda s_0)}{\cosh^2(\lambda s_0) + (\lambda s_0)^2}\ .
\label{param}
\end{equation}
Then, invoking Eqs.~(\ref{ux}), (\ref{uz}), (\ref{bihar_solution}) and (\ref{param}) the vertical displacement $\delta(x,t)=u_z(x,0,t)$ of the liquid-elastic interface reads in Fourier space:
\begin{equation}
\label{Vert_Disp}
\tilde{\delta} (\lambda,t)= \frac{1}{i \lambda} \left(\tilde{u}_x' - \frac{ i \lambda \tilde{\phi}'}{\mu}\right)(\lambda,0,t) = -\frac{\tilde{p}}{2 \mu \lambda} \, \frac{\sinh(\lambda s_0)\cosh(\lambda s_0) - \lambda s_0}{\cosh^2(\lambda s_0) + (\lambda s_0)^2}\ .
\end{equation}
Using Eqs.~(\ref{del}) and (\ref{Vert_Disp}), we find: 
\begin{equation}
\label{k}
\tilde{k}(\lambda) =  \frac{1}{2 \lambda} \, \frac{\sinh(\lambda s_0)\cosh(\lambda s_0) - \lambda s_0}{\cosh^2(\lambda s_0) + (\lambda s_0)^2}\ .
\end{equation}
In exactly the same way, the horizontal displacement $u_{\textrm{s}}(x,t)=u_x(x,0,t)$ of the liquid-elastic interface reads in Fourier space:
\begin{equation}
\tilde{u_{\textrm{s}}}(\lambda,t) =i \frac{\lambda^2 \tilde{\phi} + \tilde{\phi}''}{4 \mu \lambda} (\lambda,0,t) = \frac{i\tilde{p}}{2\mu} \, \frac{\lambda s_0^2}{\cosh^2(\lambda s_0) + (\lambda s_0)^2}\ ,
\end{equation}
which gives:
\begin{equation}
\tilde{k_{\textrm{s}}} (\lambda) =  \frac{1}{2 i} \, \frac{\lambda s_0^2}{\cosh^2(\lambda s_0) + (\lambda s_0)^2}\ .
\label{k_s}
\end{equation}

\section{Stokes-elastic model}
\label{Stokes}
The previous lubrication-elastic model assumes that the typical vertical length scale of the flow is much smaller than the horizontal one. However, the initial stepped interface and thus the early-time profiles are not compatible with this criterion. Therefore, we now instead solve the incompressible Stokes' equations for the liquid layer, in order to go beyond the lubrication approximation. 

\subsection{Hydrodynamic description of the liquid layer}
We introduce the 2D stream function $\psi$ that is related to the velocity field $\vec{v}$ via the relation $\vec{v} = \vec{\nabla} \times (\psi \vec{e_y})$ with $\vec{e_y}$ the out-of-plane unit vector and $\vec{\nabla} \times .$ the curl operator. Similarly to the Airy stress function, the stream function verifies a biharmonic equation. The kinematic and no-slip conditions at the liquid-elastic interface (located at $z=0$ at zeroth order in the perturbation, see Fig.~1(a)) imply, respectively:
\begin{subequations}
\label{BC_F}
\begin{equation}\label{NoSlip1}
v_z (x,0,t) = \partial_x \psi (x,0,t) = \partial_t u_z (x,0,t) = \partial_t \delta(x,t)\ ,
\end{equation}
\begin{equation}\label{NoSlip2} 
v_x (x,0,t) = - \partial_z \psi (x,0,t)= \partial_t u_x (x,0,t) = \partial_t u_{\textrm{s}}(x,t)\ .
\end{equation}
\end{subequations} 
In addition, at the air-liquid interface (located at $z=h_0$ at zeroth order in the perturbation, see Figs.~1(a) and~\ref{scheme}), we assume no shear and the pressure is set by the Laplace pressure. The continuity of stress thus gives:
\begin{subequations}
\label{BC_f}
\begin{equation}\label{NoShear}
\sigma_{xz} (x,h_0,t) = \eta (\partial_{xx} \psi - \partial_{zz} \psi)(x,h_0,t)=0\ ,
\end{equation}
\begin{equation}\label{Air/Liq} \begin{split} 
\sigma_{zz} (x,h_0,t) & = - \mathcal{P} (x,h_0,t) + 2 \eta  \partial_{z}(\partial_x \psi) (x,h_0,t)=-p(x,t) \ .
\end{split}
\end{equation}
\end{subequations}
with $\mathcal{P}(x,z,t)$ the excess pressure (with respect to the atmospheric pressure) in the liquid, and $p(x,t) = -\gamma \partial_{xx} h$ the Laplace pressure. Note that we neglect all nonlinear terms in $\partial_x h$ that come from the curvature in the Laplace pressure and the projection of the normal and tangential vectors onto the $x$ and $z$ axes. Now, we employ a similar method as the one developed in the previous lubrication-elastic model, and first take the Fourier transform of the biharmonic equation satisfied by the stream function:
\begin{equation}
\lambda^4 \tilde{\psi} - 2\lambda^2 \tilde{\psi}'' + \tilde{\psi}'''' = 0\ ,
\end{equation}
whose general solution is:
\begin{equation}
\label{psi}
\tilde{\psi}(\lambda,z,t)  = A_2(\lambda,t) \cosh(\lambda z) + B_2(\lambda,t) \sinh(\lambda z) + C_2(\lambda,t) z \cosh(\lambda z) + D_2(\lambda,t) z \sinh(\lambda z)\ .
\end{equation}
Taking the Fourier transforms of the boundary conditions (Eqs.~(\ref{BC_F}) and (\ref{BC_f})), we find:
\begin{subequations}
\label{psi_coeff}
\begin{equation}
- i \lambda A_2 = \partial_t \tilde{\delta}\ .
\end{equation}
\begin{equation}
\lambda B_2 = -\frac{i \tilde{p}}{2 \eta \lambda}  \, \frac{\sinh(\lambda h_0) \lambda h_0 +\cosh(\lambda h_0)}{\cosh^2(\lambda h_0) + (\lambda h_0)^2} -i \partial_t \tilde{\delta}\frac{\sinh(\lambda h_0) \cosh(\lambda h_0) - \lambda h_0}{\cosh^2(\lambda h_0) + (\lambda h_0)^2} -\partial_t \tilde{u_{\textrm{s}}} \frac{(\lambda h_0)^2}{\cosh^2(\lambda h_0) + (\lambda h_0)^2}\ ,
\end{equation}
\begin{equation}
\label{C_2}
C_2 = \frac{i \tilde{p}}{2 \eta \lambda}  \, \frac{\sinh(\lambda h_0) \lambda h_0 +\cosh(\lambda h_0)}{\cosh^2(\lambda h_0) + (\lambda h_0)^2} +i \partial_t \tilde{\delta}\frac{\sinh(\lambda h_0) \cosh(\lambda h_0) - \lambda h_0}{\cosh^2(\lambda h_0) + (\lambda h_0)^2} -\partial_t \tilde{u_{\textrm{s}}} \frac{\cosh^2(\lambda h_0)}{\cosh^2(\lambda h_0) + (\lambda h_0)^2}\ ,
\end{equation}
\begin{equation}
\label{D_2}
D_2 = \frac{- i h_0 \tilde{p}}{2 \eta} \, \frac{\cosh(\lambda h_0)}{\cosh^2(\lambda h_0) + (\lambda h_0)^2} -i \partial_t \tilde{\delta}\frac{\cosh^2(\lambda h_0)}{\cosh^2(\lambda h_0) + (\lambda h_0)^2} +\partial_t \tilde{u_{\textrm{s}}} \frac{\lambda h_0 + \sinh(\lambda h_0) \cosh(\lambda h_0)}{\cosh^2(\lambda h_0) + (\lambda h_0)^2}\ .
\end{equation}
\end{subequations}
Finally, we note that the pressure $\mathcal{P}(x,z,t)$ is entirely determined by the stream function. Indeed, in Fourier space, and invoking the stream function, the $x$-projection of the Stokes' equation reads:
\begin{equation}
i\lambda\tilde{\mathcal{P}}=\eta\left(\tilde{\psi}'''-\lambda^2\tilde{\psi}'\right)\ .
\label{pressure}
\end{equation}

\subsection{Coupling with the elastic layer}
As in the previous lubrication-elastic model, we solve the elastic part of the problem by introducing the Airy stress function $\phi$ given by Eq.~(\ref{bihar_solution}) in Fourier space. Assuming no displacement at the interface between the elastic layer and the rigid substrate (located at $z = -s_0$, see Fig.~1(a)), one has: 
\begin{subequations}
\label{BCs}
\begin{equation}
\label{BC1}
u_x (x,-s_0,t) = 0\ ,
\end{equation}
\begin{equation}
\label{BC2}
u_z (x,-s_0,t) = 0\ .
\end{equation}
\end{subequations}
Equation~(\ref{Disp_Airy}) can be used to relate the boundary conditions (Eq.~(\ref{BCs})) to the parameters $A$, $B$, $C$, $D$ (Eq.~(\ref{bihar_solution})). After some algebra, one finds: 
\begin{subequations}
\begin{equation}
\lambda A = 2 \mu \, \, \frac{ i \tilde{u_{\textrm{s}}} (\lambda s_0)^2 - \tilde{\delta} [\cosh(\lambda s_0)\sinh(\lambda s_0) + \lambda s_0]  }{ \sinh^2(\lambda s_0 ) - (\lambda s_0)^2}\ ,
\end{equation}
\begin{equation}
\lambda B = -2 \mu \tilde{\delta}\ ,
\end{equation}
\begin{equation}
C = 2 \mu \, \, \frac{- i \tilde{u_{\textrm{s}}}[\cosh(\lambda s_0) \sinh(\lambda s_0) - \lambda s_0] + \sinh^2(\lambda s_0) \tilde{\delta}}{\sinh^2(\lambda s_0 ) - (\lambda s_0)^2}\ ,
\end{equation}
\begin{equation}
D =  2 \mu \, \, \frac{ - i \tilde{u_{\textrm{s}}} \sinh^2(\lambda s_0) + \tilde{\delta} [\cosh(\lambda s_0) \sinh(\lambda s_0)  + \lambda s_0]  }{ \sinh^2(\lambda s_0 ) - (\lambda s_0)^2}\ .
\end{equation}
\end{subequations}
At the liquid-elastic interface (located at $z=0$ at zeroth order in the perturbation, see Fig.~1(a)), the normal-stress continuity reads:
\begin{equation}  
- \mathcal{P} (x,0,t)+ 2 \eta \partial_z (\partial_x \psi)(x,0,t) = \partial_{xx} \phi(x,0,t)\ ,
\end{equation}
or, equivalently, in Fourier space:
\begin{equation}
\tilde{\mathcal{P}}(\lambda,0,t)+2i\lambda\eta\tilde{\psi}'(\lambda,0,t)=\lambda^2\tilde{\phi}(\lambda,0,t)\ .
\label{ftnsc}
\end{equation}
Then, by taking the $z\rightarrow0$ limit of Eq.~(\ref{pressure}) and by combining it with Eqs.~(\ref{bihar_solution}), (\ref{psi}), and (\ref{ftnsc}), one obtains:
\begin{equation}
-2 i \eta \lambda^2 B_2 = -\lambda^2 A\ .
\label{AB2}
\end{equation}
Invoking Eqs.~(\ref{param}) and (\ref{psi_coeff}), Eq.~(\ref{AB2}) becomes:
\begin{equation}
\begin{split}
& - \tilde{p} \, \, \frac{ \cosh(\lambda h_0) + \lambda h_0
\sinh(\lambda h_0) }{\cosh^2(\lambda h_0) + (\lambda h_0)^2} -2 \eta
\lambda \partial_t \tilde{\delta} \frac{ \cosh(\lambda h_0)
\sinh(\lambda h_0) - \lambda h_0}{\cosh^2(\lambda h_0) + (\lambda h_0)^2}
+ 2 i \eta \lambda \partial_t \tilde{u_\textrm{s}}  \frac{(\lambda
h_0)^2 }{\cosh^2(\lambda h_0) + (\lambda h_0)^2}
\\
&=2 \lambda \mu \frac{ -i \tilde{u_\textrm{s}} (\lambda s_0)^2 +
\tilde{\delta} [ \cosh(\lambda s_0)\sinh(\lambda s_0) + \lambda
s_0 ] }{ \sinh^2 (\lambda s_0 ) - (\lambda s_0)^2}\ .
\end{split}
\end{equation}
For simplicity, we neglect the terms of order $T \partial_t \tilde{\delta}$ or $T \partial_t \tilde{u_{\textrm{s}}}$ with respect to the terms of order $\tilde{\delta}$ or $\tilde{u_{\textrm{s}}}$, where $T=\eta/\mu$ is a composite Maxwell-like viscoelastic time. This assumption essentially means that the elastic layer has an instantaneous response to the applied stress, or that we decouple the fast and slow dynamics and focus on the latter. This is relevant in our case since $T$ is much smaller than the experimental time scale (see inset of Fig.~4). Doing so, we get in Fourier space:
\begin{equation}
\label{Normal_stress}
- \tilde{p} \, \, \frac{ \cosh(\lambda h_0) + \lambda h_0 \sinh(\lambda h_0) }{\cosh^2(\lambda h_0) + (\lambda h_0)^2} = 2 \lambda \mu \frac{ -i \tilde{u_{\textrm{s}}} (\lambda s_0)^2 + \tilde{\delta} [ \cosh(\lambda s_0)\sinh(\lambda s_0) + \lambda s_0 ]}{ \sinh^2 (\lambda s_0 ) - (\lambda s_0)^2}\ .
\end{equation}
Besides, the tangential-stress continuity reads:
\begin{equation}
\eta (\partial_{xx} \psi - \partial_{zz} \psi ) (x,0,t)= - \partial_{xz} \phi(x,0,t)\ ,
\end{equation}
and thus, with a similar treatment, one gets in Fourier space:
\begin{equation}
\label{Tangential_stress}
\tilde{p} \, \, \frac{\lambda h_0 \cosh(\lambda h_0)}{\cosh^2(\lambda h_0) + (\lambda h_0)^2} = 2 \lambda \mu \frac{ -i \tilde{u_{\textrm{s}}} (\cosh(\lambda s_0)\sinh(\lambda s_0) - \lambda s_0 )  + \tilde{\delta} (\lambda s_0)^2  }{ \sinh^2(\lambda s_0 ) - (\lambda s_0)^2}\ .
\end{equation}  
By analogy with Eqs.~(\ref{Conv1}) and~(\ref{Conv2}) of the previous lubrication-elastic model, we introduce two new Green's functions $k_2(x)$ and $k_{\textrm{s}2}(x)$. Equations~(\ref{Normal_stress}) and (\ref{Tangential_stress}) thus lead to: 
\begin{equation}
\label{delt}
\tilde{\delta} = -\frac{\tilde{p} \tilde{k_2}}{\mu} =  \frac{- \tilde{p}}{2 \mu \lambda} \, \frac{ (\lambda s_0)^2 (\lambda h_0) \cosh(\lambda h_0)  + [ \sinh(\lambda h_0)\lambda h_0 + \cosh(\lambda h_0) ] [ \cosh(\lambda s_0) \sinh(\lambda s_0) - \lambda s_0 ] }{[\cosh^2(\lambda h_0) +(\lambda h_0)^2 ] [ \cosh^2(\lambda s_0) + (\lambda s_0)^2 ] }\ ,
\end{equation}
\begin{equation}
\label{us}
\tilde{u_{\textrm{s}}} =  -\frac{\tilde{p} \tilde{k_{\textrm{s}2}}}{\mu} = \frac{i\tilde{p}}{2 \mu \lambda} \, \frac{ \lambda h_0 \cosh(\lambda h_0) [ \cosh(\lambda s_0) \sinh(\lambda s_0) + \lambda s_0 ] + [ \cosh(\lambda h_0) + \sinh(\lambda h_0) \lambda h_0 ] (\lambda s_0)^2 }{[\cosh^2(\lambda h_0) +(\lambda h_0)^2 ] [ \cosh^2(\lambda s_0) + (\lambda s_0)^2 ] }\ ,
\end{equation}
with: 
\begin{equation}
\tilde{k_2}(\lambda) =  \frac{1}{2  \lambda} \, \frac{ (\lambda s_0)^2 (\lambda h_0) \cosh(\lambda h_0)  + [ \sinh(\lambda h_0)\lambda h_0 + \cosh(\lambda h_0) ] [ \cosh(\lambda s_0) \sinh(\lambda s_0) - \lambda s_0 ] }{[\cosh^2(\lambda h_0) +(\lambda h_0)^2 ] [ \cosh^2(\lambda s_0) + (\lambda s_0)^2 ] }\ ,
\end{equation}
\begin{equation}
\tilde{k_{\textrm{s}2}}(\lambda) =  \frac{1}{2 i  \lambda} \, \frac{ \lambda h_0 \cosh(\lambda h_0) [ \cosh(\lambda s_0) \sinh(\lambda s_0) + \lambda s_0 ] + [ \cosh(\lambda h_0) + \sinh(\lambda h_0) \lambda h_0 ] (\lambda s_0)^2 }{[\cosh^2(\lambda h_0) +(\lambda h_0)^2 ] [ \cosh^2(\lambda s_0) + (\lambda s_0)^2 ] }\ .
\end{equation}
The two Green's functions $k_2$ and $k_{\textrm{s}2}$ have forms that are quite similar to the ones of the previous lubrication-elastic model, $k$ and $k_{\textrm{s}}$ (see Eqs.~(\ref{k}) and (\ref{k_s})). Moreover, in the lubrication limit where $\lambda h_0 \to 0$, $k_2$ and $k_{\textrm{s}2}$ tend towards $k$ and $k_{\textrm{s}}$, respectively.

\subsection{Temporal evolution of the air-liquid interface}
Let us write the mass conservation for the liquid layer:
\begin{equation}
\partial_t \Delta = - \partial_x \int_{\delta(x,t)}^{h(x,t)} \mathrm{d}z\, v_x(x,z,t) = \partial_x \int_{\delta(x,t)}^{h(x,t)} \mathrm{d}z\, \partial_z \psi (x,z,t)=\partial_x\psi [x,h(x,t),t]-\partial_x\psi [x,\delta(x,t),t]\ , 
\end{equation}
with $\Delta(x,t) = h(x,t) - \delta(x,t) -h_0$ as in the previous lubrication-elastic model. At the lowest order in the perturbation, this general expression becomes:
\begin{equation}
\partial_t \Delta =\partial_x\psi (x,h_0,t)-\partial_x\psi (x,0,t)\ ,
\end{equation}
or, equivalently, in Fourier space:
\begin{equation}
\partial_t \tilde{\Delta } + i \lambda [ \tilde{\psi}(\lambda,h_0,t) - \tilde{\psi}(\lambda,0,t) ] = 0\ .
\label{mass}
\end{equation}
Using Eqs.~(\ref{Laplace}), (\ref{delt}), and~(\ref{us}), one gets:
\begin{equation}
\tilde{\delta} = \frac{-\tilde{k_2} \gamma \lambda^2}{\mu (1 + \tilde{k_2} \gamma \lambda^2/ \mu)} \tilde{\Delta}\ ,
\label{delDel}
\end{equation}
\begin{equation}
\tilde{u_{\textrm{s}}} = \frac{-\tilde{k_{\textrm{s}2}} \gamma \lambda^2}{\mu (1 + \tilde{k_2} \gamma \lambda^2 /\mu) } \tilde{\Delta}\ .
\end{equation}
By injecting Eqs.~(\ref{psi}) and~(\ref{psi_coeff}) in Eq.~(\ref{mass}), one obtains the ordinary differential equation:
\begin{equation}
\partial_t \tilde{\Delta} = -\Omega_2(\lambda) \tilde{\Delta}\ ,
\end{equation}
with:
\begin{equation}
\Omega_2(\lambda) = \frac{\gamma \lambda }{2 \eta}  \, \, \frac{\mathcal{A}(\lambda h_0)}{\mathcal{B}(\lambda h_0) + \frac{\gamma \lambda^2}{\mu } \mathcal{C}(\lambda h_0) }\ ,
\end{equation}
and:
\begin{subequations}
\begin{equation}
\mathcal{A} (\lambda h_0) = \cosh(\lambda h_0) \sinh(\lambda h_0) - \lambda h_0\ ,
\end{equation}
\begin{equation}
\mathcal{B}(\lambda h_0) = \cosh^2(\lambda h_0) + (\lambda h_0)^2\ ,
\end{equation}
\begin{equation}
\mathcal{C}(\lambda h_0) = \tilde{k_2} \left[ \cosh(\lambda h_0) +
(\lambda h_0) \sinh(\lambda h_0) \right] +i \tilde{k_{\textrm{s}2}} 
\lambda h_0 \cosh(\lambda h_0)\ .
\end{equation}
\end{subequations}
This differential equation can be solved with the initial condition (step of height $h_2$, see Fig.~1(a)):
\begin{equation}
\tilde{\Delta}(\lambda,0) = - \frac{h_2}{2i\lambda} \sqrt{\frac{2}{\pi}}\ ,
\end{equation}
thus leading to:
\begin{equation}
\tilde{\Delta}(\lambda, t) = - \frac{h_2}{2i\lambda} \sqrt{\frac{2}{\pi}} \exp\left[-\Omega_2(\lambda) t\right]\ .
\end{equation}
Then, using Eq.~(\ref{delDel}), one has:
\begin{equation}
\tilde{\Delta} + \tilde{\delta} = \frac{\tilde{\Delta}}{1 + \gamma \lambda^2 \tilde{k_2}/ \mu}\ .
\label{final}
\end{equation}
Finally, the displacement $h(x,t)-h_0=\Delta(x,t)+\delta(x,t)$ of the air-liquid interface with respect to its equilibrium position can be obtained by taking the inverse Fourier transform of Eq.~(\ref{final}).
\begin{figure}[h!]
\begin{center}
\includegraphics[scale=0.5]{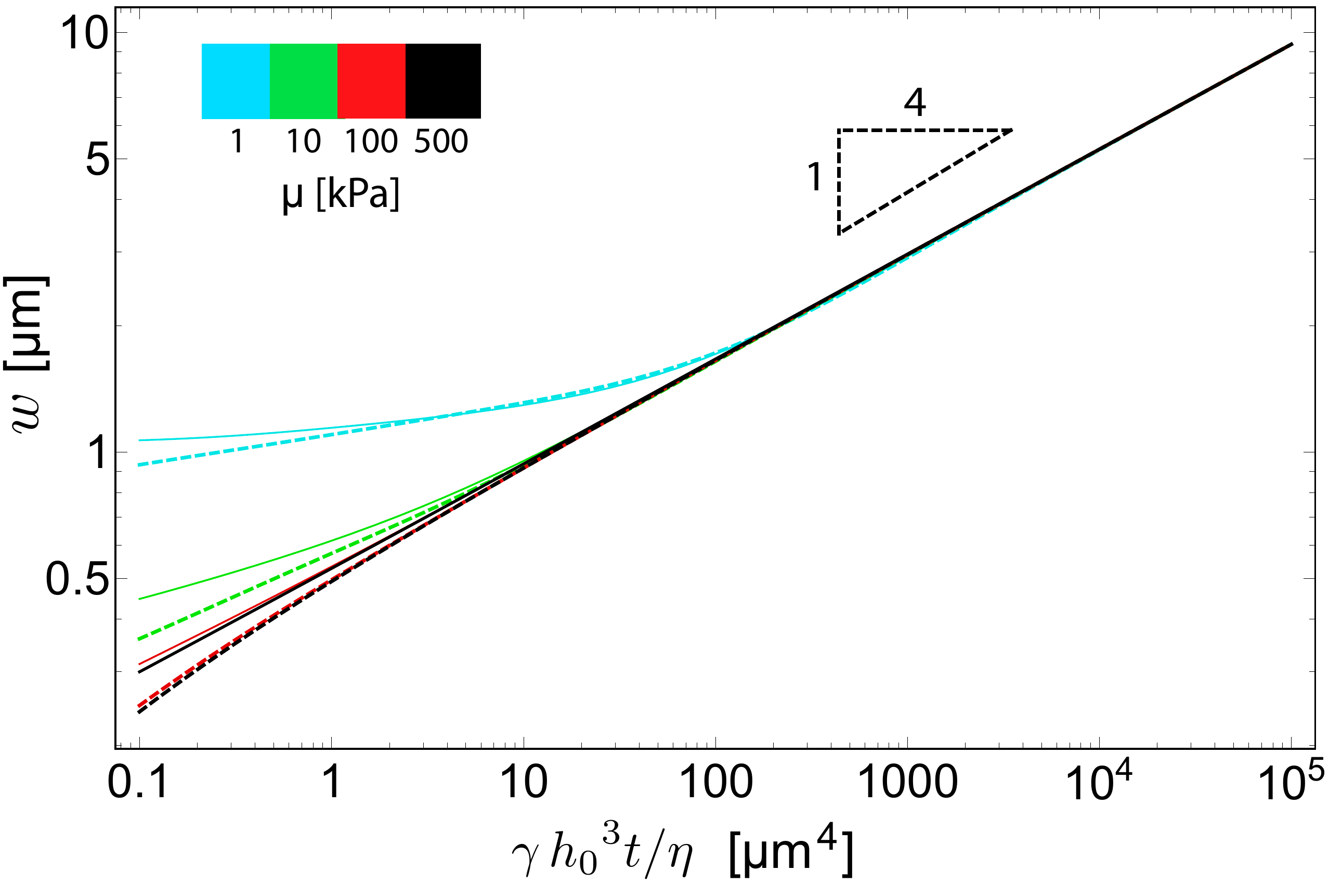}
\caption{Temporal evolution of the profile width (defined in the inset of Fig. 2(a)), in log-log scale, as predicted by both theoretical models, for different shear moduli, viscosities and liquid-film thicknesses. The $1/4$ power law corresponding to a rigid substrate is indicated. The solid lines represent the lubrication-elastic model, and the dashed lines represent the Stokes-elastic model. The shear moduli are given by the color code, which is identical to the one in Fig. 6. All the other parameters are identical to the ones used in Fig. 6.}
\label{numerical}
\end{center}
\end{figure}
Figure~\ref{numerical} displays the temporal evolutions of the profile width $\omega$ (see definition in the inset of Fig.2(a)), as derived from the two models presented in this supplementary material. 
The Stokes-elastic model exhibits the same qualitative features as the lubrication-elastic one. In particular, the width of the profile depends on elasticity only at early times, and rapidly tends to a $1/4$ power law -- characteristic of the rigid-substrate case. This result suggests that the lubrication approximation, which is not valid at early times, is not responsible for the discrepancy between the lubrication-elastic model and the experiments reported in the main text.
\end{document}